\documentclass{iopart}
\usepackage{euscript,iopams,amssymb,amsfonts,graphicx,bm}
\usepackage{pgfplots,setstack}

\usepackage{float}
\usepackage{epsfig}
\usepackage{esint}
\usepackage{chemfig}
\usepackage{cite}

\usepackage{bm,braket}

\newcommand{\R}{{\mathbb R}}

  \newcommand{\n}{\mathbf{n}}
\newcommand{\x}{\mathbf{x}}

\newcommand{\y}{\mathbf{y}}
\newcommand{\X}{\bm{X}}
\newcommand{\W}{\bm{W}}
\newcommand{\h}{{\bm h}}
\newcommand{\uhat}{\widehat{u}}
\newcommand{\E}{\mathbb E}
\newcommand{\calM}{\mathcal{M}}
\newcommand{\calU}{\mathcal{U}}
\newcommand{\calT}{\mathcal{T}}
\newcommand{\calA}{\mathcal{A}}

\newcommand{\calC}{\mathcal{C}}
\newcommand{\calP}{\mathcal{P}}

\newcommand{\calB}{\mathcal{B}}

\newcommand{\calR}{\mathcal{R}}
\newcommand{\frJ}{\mathfrak{J}}
\newcommand{\frK}{\mathfrak{K}}
\newcommand{\calJ}{\mathcal{J}}

\newcommand{\calN}{\mathcal{N}}

\eqnobysec

\newcommand{\owp}{\overline{\wp}}

\newcommand{\wlam}{\widehat{\lambda}}

\newcommand{\calS}{\mathcal{S}}
\newcommand{\calG}{{\mathcal{G}}}

\newcommand{\z}{{\bf z}}
\renewcommand{\e}{{\rm e}}

\newcommand{\hxi}{\bm \xi}

\renewcommand{\P}{\mathbb P}

\begin{document}

\title[MFT and the absorption of Brownian particles by partially reactive targets ]{Macroscopic fluctuation theory and the absorption of Brownian particles by partially reactive targets}

\author{Paul C. Bressloff}
\address{Department of Mathematics, Imperial College London, London SW7 2AZ, UK}
\ead{p.bressloff@imperial.ac.uk}

\begin{abstract}
We use macroscopic fluctuation theory (MFT) to analyse current fluctuations in a non-interacting Brownian gas with one or more partially absorbing targets within a bounded domain $\Omega \subset \R^d$. We proceed by coarse-graining a generalised Dean-Kawasaki equation with Robin boundary conditions at the target surfaces. The exterior surface $\partial \Omega$ is maintained at a constant density $\owp$ by contact with a particle bath.
We first derive Euler-Lagrange or MFT equations for the optimal noise-induced path for a single target under a saddle-point approximation of the associated path integral action. We then obtain the Gaussian distribution characterising small current fluctuations by linearising the MFT equations about the corresponding deterministic or noise-averaged system and solving the resulting stationary equations. The Robin boundary conditions are handled using the spectrum of a Dirichlet-to-Neumann operator defined on the target surface. We illustrate the theory by considering two simple geometric configurations, namely,  the finite interval and a circular annulus. In both cases we determine how the variance of the current depends on the rate of absorption (reactivity constant) $\kappa$. Finally, we extend our analysis to multiple partially absorbing targets. First, we obtain the general result that, in the case of partially absorbing targets ($0<\kappa<\infty$), the covariance matrix for current fluctuations supports cross correlations even in the absence of particle interactions. (These cross-correlations vanish in the totally absorbing limit $\kappa\rightarrow \infty$.) We then explicitly calculate the covariance matrix for circular targets in a 2D domain by assuming that the targets are much smaller than the characteristic size $L$ of the domain $\Omega$ and applying methods from singular perturbation theory.  This yields a non-perturbative expression for the covariance matrix with respect to the small parameter $\nu=-\ln \epsilon$, assuming that the target radii are $O(\epsilon L)$ with $0<\epsilon \ll 1$. The leading order contribution is a diagonal matrix whose entries are consistent with the result for a circular annulus, whereas higher-order terms generate cross-correlations.

 \end{abstract}

\maketitle    

  \section{Introduction} 
 Many processes in cell biology involve the diffusion of Brownian particles in a bounded domain $\Omega$ that contains a target $\calU$ (or possibly many targets) whose boundary $\partial \calU$ is a chemically reactive surface \cite{Bressloff22B}. The domain $\Omega$ could represent the cytoplasm of a cell and $\calU$ an intracellular compartment such as the cell nucleus or the endoplasmic reticulum. (Alternatively, the reactive target could be located within the cell membrane $\partial \Omega$ rather than the interior of the cell.) At a larger length scale, $\Omega$ could represent an extracellular volume and $\calU$ a whole cell. In this case, the diffusing particles correspond to signalling molecules that bind to the cell surface $\partial \calU$. A typical quantity of interest in these applications is the rate at which diffusing particles in the bulk react with a given target $\calU$. Mathematically speaking, the reaction rate can be identified with the total particle current at the boundary $\partial \calU$. In an unbounded domain $\R^d$ with a constant far-field density, a stationary current only exists for $d\geq 3$, reflecting the transient nature of higher-dimensional random walks, and can be determined using classical Smoluchowski reaction-rate theory \cite{Smoluchowski17,Collins49,Rice85,Redner01}.
On the other hand, a non-equilibrium stationary state (NESS) can occur irrespective of the spatial dimension in the case of a bounded domain $\Omega$, given a fixed time-independent flux or particle density on the exterior boundary $\partial \Omega$.
  
At the macroscopic level, diffusion-mediated absorption at a chemically reactive target surface $\partial \calU$ is typically modelled in terms of a diffusion equation with a Robin boundary condition on $\partial U$:
\begin{eqnarray}
\label{RB1}
&\frac{\partial \rho(\x,t)}{\partial t}=D{\bf \nabla}^2 \rho,\ \x\in \R^d \backslash \calU,\\
&-D\nabla \rho(\x,t)\cdot \n_0(\x)=\kappa \rho(\x,t),\  \x \in \partial \calU,
\label{RB2}
\end{eqnarray}
where $\rho(\x,t)$ is the particle concentration, $D$ is the diffusivity, $\n_0(\x)$ is the unit normal at points on the boundary $\partial U$, and $\kappa$ (in units of speed) is known as the reactivity constant. (For the sake of illustration, we have taken $\Omega$ to be $\R^d$.) The target surface is partially absorbing in the sense that at the single-particle level there is a nonzero probability that a particle is reflected at the surface and returns to the bulk before a reaction occurs. Note that the standard Dirichlet (totally absorbing) and Neumann (totally reflecting)  boundary conditions are recovered in the limits $\kappa\rightarrow \infty$ and $\kappa\rightarrow 0$, respectively. A non-trivial problem, even in the absence of particle-particle interactions, is relating the macroscopic description of diffusion-mediated absorption by a partially reactive target to the microscopic behaviour of individual Brownian particles and the associated current and density fluctuations. In order to understand the underlying challenges, which are the major focus of the current paper,  it is useful to consider the simpler problem of a non-interacting Brownian gas in $\R^d$ (no targets).

Suppose that there are $N_0$ independent Brownian particles in $\R^d$ and let $\X_j(t)$ denote the position of the $j$th particle. In the case of pure Brownian motion we have the stochastic differential equation (SDE) $\dot{\X}_j(t)=\sqrt{2D}{\bm \xi}_j(t)$ where ${\bm \xi}_j(t)$ is a $d$-dimensional vector of independent Gaussian white noise processes. Suppose that $\X_j(0)=\x_{j,0}$ and consider the probability density $p_j(\x,t)=\langle \delta(\X_j(t)-\x)\rangle$, where $\langle \cdot \rangle$ denotes averaging with respect to the Gaussian white noise. Using It\^o's lemma, it can be shown that $p_j$ evolves according to the single-particle diffusion or Fokker-Planck equation $\partial_t p_j=D{\bm \nabla}^2p_j$ with $p_j(\x,0)=\delta(\x-\x_{j,0})$. Summing over all the particles and defining $\rho(\x,t)=\sum_{j=1}^{N_0}p_j(\x,t)$, we obtain the multi-particle diffusion equation $\partial_t\rho=D{\bm \nabla}^2 \rho$ with $\rho(\x,0)=\sum_{j=1}^{N_0}\delta(\x-\x_{j,0})$. We thus recover the macroscopic diffusion equation, after eliminating any dependence of the initial condition on individual particles. This can be achieved either by taking $\x_{j,0}=\x_0$ for all $j=1,\ldots,N_0$ or by assuming that the initial positions are generated randomly and averaging with respect to the initial positions.

A crucial step in the derivation of the macroscopic diffusion equation from the microscopic dynamics of a Brownian gas in $\R^d$ is averaging the so-called empirical or global measure 
$\wp(\x,t)=\sum_{j=1}^{N_0} \delta(\X_j(t)-\x)$ with respect to the underlying Gaussian white noise processes. However, this eliminates all information regarding density and current fluctuations. One way to take into account such fluctuations
is to derive a Dean-Kawasaki (DK) equation for $\wp$ \cite{Dean96,Kawasaki98} and to use macroscopic fluctuation theory (MFT) \cite{Bertini15}.
The DK equation is an exact stochastic partial differential equation (SPDE) for $\wp$ in the distributional sense but it is highly singular \cite{Cornalba23}. MFT deals with the latter by coarse-graining the empirical density using a rescaling factor $\Lambda$ that is either proportional to the system size or to $\sqrt{T}$ where $T$ is a large observational time. (Mathematically speaking, coarse graining only make sense in the thermodynamic limit $N_0\rightarrow \infty$.) The resulting equation for $\widehat{\wp}(\x,t)=\wp (\Lambda \x,\Lambda^2 t)$ takes the form (after dropping the $\widehat{}$ on $\widehat{\rho}$)
\begin{eqnarray} 
 \frac{\partial {\wp}(\x,t)}{\partial t} &=-{\bm \nabla} \cdot  {\bm \frJ}(\x,t),\ \x\in \R^d ,
 \label{lattice1}
 \end{eqnarray}
 with the stochastic flux 
 \begin{eqnarray}
  {\bm \frJ}(x,t)&=   \sqrt{\frac{2D{\wp}(\x,t))}{\Lambda}} {\bm \eta}(\x,t) -D{\bm \nabla} {\wp}(\x,t) .\end{eqnarray}
Here ${\bm \eta}$ is a vector of independent spatiotemporal Gaussian white noise processes. MFT was originally developed 
to determine fluctuations in the non-equilibrium stationary states of
diffusive lattice gas models \cite{Bertini01,Bertini02,Bertini05,Tailleur07,Burin13,Lasinio14}, with fluxes of the more general form
\begin{equation}
\label{lattice2}
{\bm \frJ}(x,t) =   \sqrt{\frac{\sigma( {\wp}(\x,t))}{\Lambda}} {\bm \eta}(\x,t) -D( {\wp}(\x,t)){\bm \nabla} {\wp}(\x,t) .
\end{equation}
(A fluctuating lattice gas is fully characterised by its diffusivity $D(\wp)$ and mobility function $\sigma(\wp)$.)
However, MFT has subsequently been applied in various non-stationary settings
\cite{Derrida07,Derrida09,Kaprivsky12,Kaprivsky12a,Meerson13,Hurtado14,Meerson14,Vilenkin14,Kaprivsky15}, including the survival of Brownian particles in the presence of totally absorbing interior or exterior boundaries \cite{Meerson14a,Meerson14b,Meerson15,Agranov16,Agranov17,Agranov18}. 
A major advantage of coarse-graining is that the corresponding SPDE operates in a weak noise regime. This means that the path integral representation of the probability functional of sample paths $\{({\wp}(\x,t), \frJ [\x,t])\}$ can be approximated using a saddle-point method. The dominant contribution to the path integral is then determined by minimising the associated action functional, resulting in a variational principle that can be used to derive various large deviation results for density and current fluctuations.
  
The main goal of the current paper is to apply MFT to a non-interacting Brownian gas in the presence of one or more partially absorbing targets. (Possible extensions to interacting particle systems are briefly discussed in section 6.) 
The structure of the paper is as follows. In section 2 we derive a higher-dimensional version of the generalised DK equation introduced in Ref. \cite{Bressloff24}. At the microscopic level this is based on a multi-particle version of the stochastic Skorokhod equation for reflected Brownian motion \cite{Freidlin85}. Heuristically speaking, each particle is independently given an impulsive lick whenever it hits the boundary, which can be represented as the differential of the corresponding boundary local time. The latter is a Brownian functional that effectively measures the amount of particle-surface contact time \cite{Ito63,Ito65,McKean75}. In addition, each particle is independently absorbed when its local time exceeds a randomly generated (quenched) local time threshold. Taking the local time threshold to be exponentially distributed provides the probabilistic underpinning of the Robin boundary condition for the Fokker-Planck equation of a single Brownian particle \cite{Grebenkov20,Grebenkov22,Bressloff22,Bressloff22a,Grebenkov24}. Here we show that in order to obtain a closed DK equation for a corresponding Brownian gas, it is necessary to impose a mean field ansatz with respect to threshold averaging.

In section 3 we apply MFT to a coarse-grained version of the generalised DK equation for a Brownian gas in a bounded domain $\Omega\backslash \calU$ with $\calU \subset \Omega \subset \R^d$. The exterior surface $\partial \Omega$ is maintained at a constant density $\owp$ by contact with a particle bath, say, whilst the interior surface is taken to be a partially absorbing target. Extending a previous study of totally absorbing targets \cite{Agranov17} in $\R^3$, we derive Euler-Lagrange (MFT) equations for the optimal noise-induced path under a saddle-point approximation of the corresponding path integral representation. We also assume that at large times $T\gg L^2/D$, where $L$ specifies the characteristic size of the domain $\Omega\backslash \calU$, typical current fluctuations at the target surface are determined by linearising the MFT equations about the corresponding deterministic or noise-averaged system and solving the resulting stationary equations. The latter involve a mixture of homogeneous and inhomogeneous steady-state diffusion equations with Robin boundary conditions on $\partial \calU$. These are formally solved in terms of the eigenvalues and eigenfunctions of a corresponding Dirichlet-to-Neumann (D-to-N) operator on $\partial \calU$ \cite{Grebenkov20d}. We thus obtain the probability distribution for small current fluctuations
\begin{equation}
\label{calJ0}
\calJ(T)=\frac{1}{T}\int_0^T \bigg [\int_{\partial \calU} {\bm \frJ}(\y,t)\cdot \n_0(\y)d\y\bigg ]dt,
\end{equation}
which takes the general form
\begin{equation}
\label{LDP}
\calP(\calJ(T)=\calJ ,T)\asymp \exp \left (-\frac{T}{2 \calC(\kappa)}\bigg [\calJ-\calJ^*\bigg ]^2 \right ).
\end{equation}
Here $\calC(\kappa) $ is a configuration-dependent function of the reactivity $\kappa$ and $\calJ^*$ is the stationary current of the deterministic system. The variance of the current fluctuations is thus $\langle \delta \calJ^2\rangle = \calC(\kappa)/T$. We then consider two simple geometric configurations where explicit solutions can be obtained without recourse to spectral theory: (i) The interval $[0,L]$ with a partially absorbing target at $x=0$ and a fixed density at $x=L$; (ii) a circular annulus with a partially absorbing inner surface of radius $R$ and a fixed density on the outer surface of radius $L$. In particular, we derive the following explicit expressions for the variances:
\begin{eqnarray}
\fl \langle \delta \calJ^2\rangle_{\rm 1D}=\frac{\owp D}{TL}   \frac{ (1+2D/\kappa L)}{(1+D/\kappa  L)^3},\quad \langle \delta \calJ^2\rangle_{\rm 2D}=
\frac{2\pi   \owp D}{T}  \frac{  {2D}/{\kappa R} +\ln(L/R)}{ [D/\kappa R+ \ln(L/R)]^3}  \ln(L/R).\nonumber \\
\fl
\end{eqnarray}

Finally, in sections 4 and 5, we extend our analysis of small current fluctuations to the case of multiple partially absorbing targets in a bounded domain $\Omega$. First, we derive a multi--target version of equation (\ref{LDP}) for the currents
\begin{equation}
\label{nhsmulti0}
\calJ_a(T)\equiv \frac{1}{T}\int_0^T \bigg [\int_{\partial \calU_a} {\bm \frJ}(\y,t)\cdot \n_a(\y)d\y\bigg ]dt,
\end{equation}
where $\partial \calU_a$ is the surface of the $a$-th target, $a=1,\ldots,M$, and $\n_a$ is the corresponding unit normal. We show that the probability distribution has the general structure (see section 4)
\begin{equation}
\calP[\delta{\bm \calJ}(T)=\delta {\bm \calJ},T]\asymp \exp\left (-\frac{T}{2}\delta {\bm  \calJ}^{\top} {\bm \Sigma}^{-1}(\kappa) \delta {\bm \calJ}\right ).
\label{mLDP}
\end{equation}
Here $\delta {\bm \calJ}$ is the vector of current fluctuations and $  {\bm \Sigma}(\kappa) /T$ is a $\kappa$-dependent covariance matrix. One of the major differences from the case of totally absorbing targets \cite{Agranov17} is that $  {\bm \Sigma}(\kappa) $ has non-zero off-diagonal terms when $0<\kappa <\infty$. This means that there are cross-correlations in the statistics of current fluctuations even though the  Brownian gas is non-interacting. (These cross-correlations vanish in the totally absorbing limit $\kappa \rightarrow \infty$.)

Calculating the covariance matrix for a general multi-target configuration in $\Omega\subset \R^d$ with $d>1$ is considerably more involved than the single target case. Therefore, in section 5 we assume that the targets are much smaller than the domain size $L$ and apply methods from singular perturbation theory, see the recent review \cite{Bressloff24a} and references therein. The basic idea is to construct an inner or local solution in a small neighbourhood of each target. The set of inner solutions is then matched to an outer solution in the bulk domain by expanding the latter in terms of the Green's function of the diffusion equation in the absence of any targets. The details of the matched asymptotic analysis differ significantly in 2D and 3D due to corresponding differences in the Green's function singularities: 
\begin{eqnarray*}
&G_{\Omega}(\x,\x_0)\rightarrow -\frac{1}{2\pi D}\ln|\x-\x_0| \mbox{ in 2D },  \\ &G_{\Omega}(\x,\x_0)\rightarrow \frac{1}{4\pi D|\x-\x_0|} \mbox{ in 3D}
\end{eqnarray*}
as $|\x-\x_0|\rightarrow 0$.
Consequently, an asymptotic expansion of the solution to a singularly perturbed BVP is typically in powers of $\epsilon$ in 3D and $\nu=-1/\ln \epsilon$ in 2D, where $\epsilon$ represents the size of a target relative to the size of the bulk domain. For the sake of illustration, we focus on circular targets in a 2D domain. (This is also particularly interesting since classical Smoluchowski theory doesn't hold in the unbounded case $\Omega =\R^2$.) Assuming an $\epsilon$-dependent reactivity $\kappa =  \kappa_0/\epsilon$ and a diffusivity $D=D_0/\nu$, we derive 
an expression for the matrix ${\bm \Sigma}={\bm \Sigma}(\kappa_0,\nu) $ that is a non-perturbative function of the small parameter $\nu$. Expanding the latter as a power series in $\nu$ then yields
\begin{equation}
\bigg \langle \delta {\calJ}_a(T)  \delta {\calJ}_b(T)\bigg \rangle \equiv \frac{1}{T}\bigg [\Sigma_{ab}^{(0)}+\nu \Sigma_{ab}^{(1)}+\ldots
\bigg ].\label{sig0}
\end{equation}
Assuming that each target is a circle of radius $\epsilon R_a$, we find that the leading order contribution is a diagonal matrix 
\begin{equation}
\Sigma_{ab}^{(0)}(T)=\frac{2\pi \owp  D_0}{T}\frac{1+2D_0/\kappa_0 R_a}{(1+D_0/\kappa_0 R_a)^3} \delta_{a,b}.
\end{equation}
On the other hand, $\Sigma_{ab}^{(1)}$ is a non-diagonal matrix that takes into account diffusion-mediated contributions to cross-correlations between target current fluctuations.

\section{Generalised DK equation for a single partially absorbing target}

Consider $N_0$ identical, non-interacting Brownian particles diffusing in the domain $\R^d\backslash \calU$ with a partially absorbing target surface $\partial \calU$. 
The generalised DK equation is an exact SPDE for the empirical measure \cite{Bressloff24}
\begin{equation}
\label{wp}
\fl \wp(\x,\h,t) =\sum_{j=1}^{N_0} \wp_{j}(\x,h_j,t) \equiv \sum_{j=1}^{N_0} \delta(\X_j(t)-\x)\Theta(h_j-L_j(t)), \quad h_j>0,
\end{equation}
where $\Theta(x)$ is a Heaviside function, $L_j(t)$ is the boundary local time of the $j$th particle,
\begin{equation}
\label{Lj}
L_j(t)=D\int_0^t\left (\int_{\partial \calU}\delta (\X_j(\tau)-\y)d\y\right )d\tau,
\end{equation}
$h_j$ is a local time threshold with $\h=(h_1,\ldots,h_N)$, and $\X_j(t)$ evolves according to the  stochastic Skorokhod equation \cite{Freidlin85}
\begin{equation}
\label{Sko}
d\X_j(t)=\sqrt{2D}d{\bf W}_j(t)-\n_0(\X_j(t))dL_j(t),
\end{equation}
with ${\bf W}_j(t)$ a vector of independent Brownian motions. In particular $\langle W_{j,k}(t)\rangle =0$ and $\langle W_{j,k}(t)W_{j',k'}(t) \rangle=\delta_{j,j'}\delta_{k,k'}\min\{t,t'\}$ with $k,k'=1,\ldots,d$. The above equations have the following interpretations. First, the local time $L_j(t)$ keeps track of the amount of time that the $j$th Brownian particle spends in the neighbourhood of points on the target boundary $\partial \calU$.
(A common convention is to include the factor $D$ in the definition of the local time, which means that $L_j(t) $ has units of length. It can be proven that $L_j(t)$ exists and is a nondecreasing, continuous function of $t$ \cite{Ito63,Ito65,McKean75}.) Second, heuristically speaking, the differential of the local time appearing in the SDE (\ref{Sko}) generates an impulsive kick directed away from the target whenever the particle encounters the boundary. 
Third, the Heaviside function $\Theta(h_j-L_j(t))$ implies that the $j$th particle is absorbed by the target when its local time crosses a threshold $h_j$, $h_j>0$. This leads to the following definition of the stopping time or first-passage time for absorption:
\begin{equation}
\label{Tj}
{\mathcal T}_j=\inf\{t>0:\ L_j(t) >h_j\}.
\end{equation}
The local time threshold $h_j$ can be interpreted physically as conditioning the probability of absorption on the amount of particle-surface contact time as determined by the local time. We will take $h_j$ to be a quenched random variable generated from an exponential probability density $\psi(h)=z\e^{-zh}$. This is equivalent to imposing a Robin boundary condition with reactivity $\kappa= zD$ on the SPDE for the corresponding particle density, see below and Ref. \cite{Bressloff24}.

\subsection{Derivation of generalised DK equation}

We begin by deriving the multi-dimensional version of the 1D generalised Dean equation considered in Ref. \cite{Bressloff24}. Consider an arbitrary smooth test function $f(\x)$ and set
\begin{equation}
{\mathfrak F}(t)\equiv  \sum_{j=1}^{N_0} f(\X_j(t)) \Theta(h_j-L_j(t))=\int_{\R^d\backslash \calU} d\x\, \wp(\x,\h,t)f(\x).
\end{equation}
Using It\^o's lemma to evaluate $d{\mathfrak F}(t)$ yields
\begin{eqnarray}
\fl &\int_{\R^d\backslash \calU}d\x\, f(\x)\frac{\partial \wp (\x,\h,t)}{\partial t} + D\int_{\partial \calU} d\y\, f(\y) \bigg (\sum_{j=1}^{N_0}\delta(\X_j(t)-\y)  \delta(h_j-L_j(t))\bigg )\nonumber \\
\fl  &=\sqrt{2D}\int_{\R^d\backslash \calU}d\x \sum_{j=1}^{N_0} \wp_{j}(\x, h_j, t)  \bigg ({\bm \nabla}f(\x)  \cdot {\bm \xi}_j(t)\bigg )\nonumber \\
 \fl  &\quad +D\int_{\R^d\backslash \calU }d\x \, \wp(\x, \h, t)\bigg ( {\bm \nabla}^2f(\x)+\int_{\partial \calU}\delta(\x-\y){\bm \nabla }f(\y)\cdot {\bf n}_0(\y)\bigg ). \label{beav}
\end{eqnarray}
We have formally set $d\W_i(t)={\bm \xi}_i(t)dt$, where ${\bm \xi}_i(t)$ is a vector of Gaussian white-noise processes such that
\begin{equation}
\langle \xi_{i,k}(t)\rangle,\quad \langle \xi_{i,k}(t)\xi_{j,l}(t')\rangle =\delta(t-t')\delta_{i,j}\delta_{k,l}
\end{equation}
for $k,l=1,\ldots,d$.
Integrating by parts the various terms on the right-hand side of equation (\ref{beav}) gives
\begin{eqnarray}
 & \int_{\R^d\backslash \calU}d\x\, f(\x)\frac{\partial \wp(\x,\h,t)}{\partial t}   \nonumber \\
  &= \int_{\R^d\backslash \calU}d\x\,   f(\x)\left (-\sqrt{2D}\sum_{i=1}^{N_0}{\bm \nabla}\wp_{i}(\x,h_i,t)  \cdot {\bm \xi}_i(t)+D{\bm \nabla}^2 \wp(\x,\h,t)\right ) \nonumber  
\\
 &\ +\int_{\partial \calU}d\y\,  f(\y) \bigg (\sqrt{2D}\sum_{i=1}^{N_0}\wp_{i}(\y,h_i,t)\ {\bm  \xi_i}(t)\cdot \n_0(\y) \nonumber \\
 &\hspace{2cm}-D{\bm \nabla} \wp(\y,\h,t)\cdot \n_0(\y) -D\nu(\y,\h,t)\bigg ),
 \label{RHS}  \end{eqnarray}
 where we have introduced a second empirical measure
 \begin{equation}
 \label{nu}
 \nu(\x,\h,t)= \sum_{j=1}^{N_0}\delta(\X_j(t)-\x)  \delta(h_j-L_j(t)) . \end{equation}
Using the fact that $f(\x)$ is arbitrary, we obtain the SPDE
\begin{eqnarray}
\label{atoto1}
\fl \frac{\partial \wp(\x,\h,t)}{\partial t} 
=-\sqrt{2D} \sum_{i=1}^{N_0}{\bm \nabla}\wp_{i}(\x,h_i,t)  \cdot {\bm \xi}_i(t)+D{\bm \nabla}^2 \wp(\x,\h,t),\quad \x\in \R^d\backslash \calU,
\end{eqnarray}
with the boundary condition
\begin{eqnarray}
\fl &\sqrt{2D}\sum_{i=1}^{N_0} \wp_{i}(\y,h_i,t) {\bm \xi}_i(t)\cdot \n_0(\y)-D{\bm \nabla} \wp(\y,\h,t)\cdot \n_0(\y)=D\nu(\y,\h,t) ,\ \y \in \partial  \calU.
\label{atoto2}
\end{eqnarray}

We now apply a noise transformation along the lines of Ref.  \cite{Dean96}. That is, define the spatio-temporal noise
\begin{equation}
\xi_k(\x,\h,t)=\sum_{i=1}^N \delta(\X_i(t)-\x)\Theta(h_i-L_i(t))\xi_{i,k}(t)
\end{equation}
such that $\langle \xi_k(\x,\h,t)\rangle =0$ and
\begin{eqnarray*}
\fl \langle \xi_k(\x,\h,t)\xi_{k'}(\x',\h,t') \rangle &=\delta_{k,k'}\delta(t-t')\sum_{i=1}^N \delta(\X_i(t)-\x)\delta(\X_i(t)-\x')\Theta(h_i-L_i(t))^2\\
&=\delta_{k,k'}\delta(t-t')\delta(\x-\x')\sum_{i=1}^N \delta(\X_i(t)-\x)\Theta(h_i-L_i(t))\\
&=\delta_{k,k'}\delta(t-t')\delta(\x-\x')\wp(\x,\h,t).
\end{eqnarray*}
It can be checked that ${\bm \xi}(\x,\h,t)$ has the same Gaussian statistics as
\begin{equation}
\widehat{\bm \xi}(\x,\h,t)=\sqrt{\wp(\x,\h,t)}{\bm \eta}(\x,t)
\end{equation}
where ${\bm \eta}(\x,t)$ denotes a vector of independent spatiotemporal Gaussian white noise processes with
\begin{equation}
\langle  \eta_k(\x,t)\rangle =0,\quad \langle  \eta_k(\x,t)\eta_l(\y,t')\rangle =\delta_{k,l}\delta(t-t')\delta(\x-\y) ,
\end{equation}
with $k,l=1,\ldots,d$. 
Hence, we obtain the stochastic DK equation 
\numparts
\begin{eqnarray} 
\label{toto1}
 &\frac{\partial \wp(\x,\h,t)}{\partial t} =-{\bm \nabla} {\bm \frJ}(\x,\h,t) ,\quad \x\in \R^d\backslash \calU,\\
& {\bm \frJ}(\y,\h,t)\cdot \n_0(\y)= \nu(\y,\h,t),\quad \y\in \partial \calU
\label{toto2}\end{eqnarray}
\endnumparts
for
\begin{eqnarray}
\label{toto3}
 {\bm \frJ}(\x,\h,t) &=   \sqrt{2D \wp(\x,\h,t)} {\bm \eta}(\x,t) -D{\bm \nabla}\wp(\x,\h,t).
\end{eqnarray}

One of the major difficulties with equations (\ref{toto1}) and (\ref{toto2}) is that they do not correspond to a closed system of equations for $\wp(\x,\h,t)$ due to the coupling with the second
empirical measure $\nu(\x,\h,t)$ at the surface $\partial \calU$. In other words, we have a closure problem.
Following our previous work \cite{Bressloff24}, suppose that we average equations (\ref{toto1}) and (\ref{toto2}) with respect to the Gaussian white noise processes and the quenched random thresholds. (If pairwise particle interactions were included then averaging the resulting DK equation would lead to a moment closure problem \cite{Dean96}, see the discussion (section 6).)
Denoting the averaged density by $\rho^{\psi}(\x,t)$, we have
\begin{eqnarray}
\fl {\rho^{\psi}}(\x,t)&= \prod_{i=1}^{\calM}\bigg [ \int_0^{\infty} \psi(h_i) dh_i\bigg ]  \sum_{j=1}^{N_0}\bigg \langle \delta(\X_j(t)-\x) \Theta(h_j-L_j(t)\bigg \rangle\nonumber \\
\fl &
= \int_0^{\infty} dh \,  \Psi(h)\mu_h(\x,t),
\label{EB2}
\end{eqnarray}
where $\Psi(h)=\int_h^{\infty} \psi(h')dh'$ and
\begin{equation}
\label{ltp}
\mu_h(\x,t)=\sum_{j=1}^{N_0}\bigg \langle \delta(\X_j(t)-\x)\delta(h-L_j(t))\bigg \rangle.
\end{equation}
Throughout the paper, angular brackets are used to denote expectations with respect to the Gaussian white noise processes only.
We have used the fact that the local time thresholds $h_j$ are generated from the same probability distribution.
Taking double expectations of equations (\ref{toto1}) and (\ref{toto2}) then yields the deterministic  BVP
\numparts
\begin{eqnarray} 
\label{Eprop1}
 &\frac{\partial \rho^{\psi}(\x,t)}{\partial t} =D{\bm \nabla}^2 \rho^{\psi}(\x,t),\quad \x\in \R^d\backslash \calU\\
 \label{Eprop2}
&{\bm \nabla}  \rho^{\psi}(\y,t)\cdot \n_0(\y)= -\mu^{\psi}(\y,t),\quad \y \in \partial \calU,\end{eqnarray}
\endnumparts
where
\begin{eqnarray}
\label{mu}
 {\mu^{\psi}}(\x,t)\equiv \int_0dh\, \psi(h) \mu_h(\x,t).
\end{eqnarray}
Both $\rho^{\psi}(\x,t)$ and $\mu^{\psi}(\x,t)$ involve weighted integrals of the density $\mu_h(\x,t)$. The latter is a multi-particle version of the so-called local time propagator \cite{Grebenkov20,Bressloff24}, which is the joint probability density for the particle positions and local times averaged with respect to the Gaussian white noise processes.

As it stands, the system of equations (\ref{Eprop1})--(\ref{Eprop1}) still does not form a closed BVP for $\rho^{\psi}$ since it couples to $\mu^{\psi}$. However, in the specific case of an exponentially distributed threshold, $\psi(h)=z\e^{-zh}$ and $\Psi(h)=\e^{-zh}$, we have $\rho^{\psi} = \rho$ and $\mu^{\psi}=z\rho$ where $\rho$ is the solution of the Robin BVP
\numparts
\begin{eqnarray} 
\label{RBP1}
 &\frac{\partial \rho(\x,t)}{\partial t} = D{\bm \nabla}^2  \rho(\x,t),\quad \x\in \R^d\backslash \calU,\\
& {\bm \nabla}  \rho(\y,t)\cdot \n_0(\y)= -\kappa \rho(\y,t),\quad \y \in \partial \calU .
\label{RBP2}
\end{eqnarray}
\endnumparts
 The effective reactivity constant is $\kappa=zD$. The mapping between an exponentially distributed local time threshold at the SDE level to the Robin BVP at the PDE level is a well-established result in probability theory \cite{Freidlin85,Papanicolaou90,Grebenkov06,Grebenkov07,Singer08,Grebenkov19a}. 
  Moreover, equation (\ref{EB2}) implies that ${\rho}(\x,t)$ is formally identical to the Laplace transform of $\mu_h(\x,t)$ with respect to $h$. Hence, one way to calculate $\mu_h(x,t)$ is to solve the Robin BVP (\ref{RBP1})-(\ref{RBP2}) and then calculate the inverse Laplace transform with respect to $z$:
  \begin{equation}
 \mu_h(\x,t)= \mbox{LT}^{-1}_{h}[{\rho}(\x,t)] .
 \label{LT}
  \end{equation}
 One can then extend the theory to non-Markovian reactive surfaces by integrating $\mu_h(\x,t)$ with respect to a non-exponential threshold distribution $\Psi(h)$, which constitutes the encounter-based approach to diffusion-mediated surface absorption at the single-particle \cite{Grebenkov20,Grebenkov22,Bressloff22,Bressloff22a,Grebenkov24} and multi-particle \cite{Bressloff24} levels. Some physical examples of non-Markovian absorption processes can be found in Refs. \cite{Bartholomew01,Filoche08}. 
  
  \subsection{Robin boundary condition for the stochastic DK equation} Averaging with respect to the Gaussian white noise processes and the quenched local time thresholds eliminates all information about current fluctuations at the target surface. In this paper, we partially preserve such information by only averaging with respect to the quenched random thresholds using the mean field ansatz
  \begin{equation}
  \E\bigg [  \sqrt{\wp(\x,\h,t)}\bigg ] =\sqrt{ \E [  \wp(\x,\h,t) ]}.
  \label{ans}
  \end{equation}
  Here $\E[f(\h)]=\bigg [  \prod_{i=1}^{\calM}\int_0^{\infty}dh_i\,  \psi(h_i) \bigg ]f(\h)$.
   Averaging equations (\ref{toto1})--(\ref{toto3}) with respect to exponentially distributed thresholds then gives a stochastic DK equation for $\wp(\x,t)=\E[\wp(\x,
   \h,t)]$ with a Robin boundary condition on $\partial \calU$:
\numparts
\begin{eqnarray} 
\label{RBDK1}
 &\frac{\partial \wp(\x,t)}{\partial t} =-{\bm \nabla} {\bm \frJ}(\x,t) ,\quad \x\in \R^d\backslash \calU,\\
& {\bm \frJ}(\y,t)\cdot \n_0(\y)= \kappa\wp(\y,t),\quad \y\in \partial \calU
\label{RBDK2}\end{eqnarray}
\endnumparts
for
\begin{eqnarray}
\label{RBDK3}
 {\bm \frJ}(\x,t) &=   \sqrt{2D \wp(\x,t)} {\bm \eta}(\x,t) -D{\bm \nabla}\wp(\x,t)
\end{eqnarray}
and the initial condition
$\wp(\x,0)=N_0\delta(\x-\x_0)$. 

The stochastic DK equation (\ref{RBDK1}) with the Robin boundary condition (\ref{RBDK2}) is the starting point for our MFT analysis of current fluctuations at partially reactive targets. Of course, we could simply have written down equations (\ref{RBDK1}) and (\ref{RBDK2}) without deriving them from first principles. However, it is important to 
understand the relationship between the SPDE and the underlying microscopic theory. In particular, the Robin boundary condition is obtained by averaging the full DK equations (\ref{toto1}) and (\ref{toto2}) with respect to a set of exponentially distributed quenched local time thresholds and imposing the mean field ansatz (\ref{ans}). This relationship will be explored further in the discussion (section 6).

\setcounter{equation}{0}

\section{MFT of current fluctuations at a single target}

\begin{figure}[b!]
\centering
\includegraphics[width=10cm]{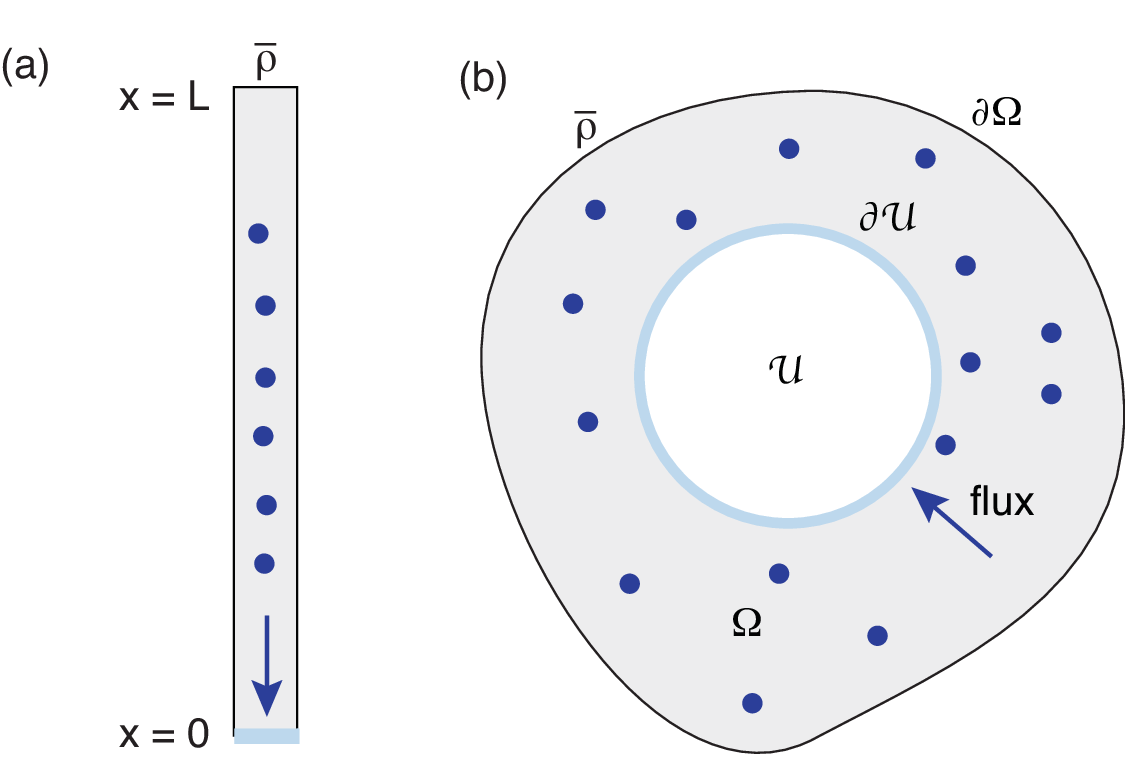} 
\caption{Brownian particles diffusing in a bounded domain $\Omega$ containing a single partially absorbing target $\calU$. A constant density $\overline{\rho}$ is maintained on the exterior boundary. (a) Finite interval $\Omega =[0,L]$ with a target at $x=0$. (b) 2D domain $\Omega$ containing a singule circular target $\calU$.}
\label{fig1}
\end{figure}

In this section we use MFT to derive a variational formulation of current fluctuations in a coarse-grained version of the DK equations (\ref{RBDK1}) and (\ref{RBDK2}). Motivated by various examples from cell biology, we take $\calU$ to be located in the interior of a bounded domain $\Omega \subset \R^d$ with a fixed density $\overline{\wp}$ on the boundary $\partial \Omega$, that is, $\wp(\x,t)=\owp$ for all $\x\in  \partial \Omega$. Examples in 1D and 2D are shown in Fig. \ref{fig1}. It follows that the averaged system supports a nonequilibrium stationary state (NESS) $\wp^*(\x)$ with a corresponding stationary flux ${\bm \frJ}^*(\x)=-D{\bm \nabla} \wp^*$. Let $\calN(T)$ denote the stochastic number of particles that are absorbed by the target over the time interval $[0,T]$ and define the corresponding rate of absorption or current
\begin{equation}
\label{calJ}
\calJ(T)=\frac{\calN(T)}{T}=\frac{1}{T}\int_0^T \bigg [\int_{\partial \calU} {\bm \frJ}(\y,t)\cdot \n_0(\y)d\y\bigg ]dt.
\end{equation}
Following along analogous lines to Ref. \cite{Agranov17}, we will derive the probability distribution (\ref{LDP}) for current fluctuations at large times $T\gg L^2/D$, where
 $L$ denotes the characteristic size of the domain, $|\Omega\backslash \calU|=L^d$.
 
\subsection{Variational formulation}

The first step in MFT is to coarse-grain equations (\ref{RBDK1}) and (\ref{RBDK2}) using the rescaling factor $\sqrt{T}$ with $T\gg L^2/D$. Under the rescaling scheme
\begin{equation}
\label{hatvar}
 \widehat{\x}=\frac{\x}{\sqrt{T}},\quad \widehat{t}=\frac{t}{T},\quad \widehat{\kappa}=\sqrt{T} \kappa,\quad \widehat{L}=\frac{L}{\sqrt{T}}
\end{equation}
we define the coarse-grained fields
\begin{equation}
q (\widehat{\x},\widehat{t})=\wp (\sqrt{T}\widehat{\x},T \widehat{t}),\quad {\bf j} (\widehat{\x},\widehat{t})=\sqrt{T}{\bm \frJ} (\sqrt{T}\widehat{\x},T \widehat{t}).
\end{equation}
The rescaling of the reaction rate $\kappa$ is consistent with the fact that it has the dimension of a velocity. The resulting coarse-grained macroscopic system (after dropping the $\widehat{}$ labels) takes the form
\numparts
\begin{eqnarray} 
\label{DKq1}
 &\frac{\partial q(\x,t)}{\partial t} =-{\bm \nabla} {\bm j}(\x,t) ,\quad \x\in \Omega\backslash \calU,\\
\label{DKq2}
&{\bm j}(\y,t)\cdot \n_0(\y)= \kappa   q(\y,t) ,\quad \y\in \partial \calU,\\
&q(\y,t)=\owp,\quad \y \in \partial \Omega,
\label{DKq3}\end{eqnarray}
\endnumparts
with the flux
\begin{eqnarray}
\label{DKq4}
&{\bm j}(\x,t) =  \sqrt{\frac{2D}{T^{d/2}}q(\x,t)} {\bm \eta}(\x,t) -D{\bm \nabla}q(\x,t),
\end{eqnarray}
and the initial condition
$q(\x,0)=\rho_0\delta(\x-
\x_0)$.
We have set $\rho_0=T^{-d/2}N_0$.
Using the definition of $\calJ(T)$, see equation (\ref{calJ}), we impose the additional constraint 
\begin{equation}
\label{conJ}
\int_0^1 \bigg [\int_{\partial \calU}{\bm j}(\y,t)\cdot \n_0(\y) d\y \bigg ]dt =T^{1-d/2}\calJ .
\end{equation}

A crucial feature of the coarse-grained model given by equations (\ref{DKq1})-(\ref{DKq3}) is that the noise is weak, which means path integral methods can be used to analyse fluctuations in the large-time limit. In particular, since the dynamics in the bulk is described by the standard SPDE for a non-interacting Brownian gas, we can use the classical Martin-Siggia-Rose formalism \cite{Martin73,Tailleur07,Derrida09,Agranov17} to obtain the following path integral representation of the probability functional for a particular realisation of $\{(q(\x,t),{\bm j}(\x,t)), \ t\in [0,T]\}$:
\begin{eqnarray}
\fl  \calP &\simeq\int D[q,{\bm j},p] \exp \left (-T^{d/2}\int_0^1 dt\, \int_{\Omega}d\x\,  \frac{[{\bm j}(\x,t)+D{\bm \nabla}q(\x,t)]^2}{4Dq(\x,t)}\right )\nonumber \\
\fl&\quad \times \exp \left (-T^{d/2}\int_0^1dt\, \int_{\Omega\backslash \calU}d\x\, p(\x,t)\left (\partial_tq(\x,t)+{\bm \nabla }\cdot {\bm j}(\x,t)\right )\right ).
\label{pathint}
\end{eqnarray}
The auxiliary field $p(\x,t)$ imposes the constraint $\partial_tq+{\bm \nabla}\cdot {\bm j}=0$. In the large-$T$ limit, the path integral can be evaluated using a saddle-point approximation.
The dominant contribution arises from the most probable path under the additional constraint (\ref{conJ}). Incorporating the latter using the Lagrange multiplier $\lambda$ implies that
\begin{equation}
\label{LDP0}
-\ln \calP[\calJ(T)=\calJ,T]\asymp T^{d/2} \calS,
\end{equation}
with the effective action
\begin{eqnarray}
\calS&=\min_{q,{\bm j},p}\bigg \{ \int_0^1 dt\, \int_{\Omega\backslash \calU}d\x\, \bigg [  \frac{[{\bm j} +D{\bm \nabla}q]^2}{4Dq} 
 +p[\partial_tq+{\bm \nabla}\cdot {\bm j}] \bigg ]\nonumber \\
 &\hspace{2cm} -\lambda  \bigg ( \int_0^1 dt \int_{\partial \calU}d\y \, {\bm j}\cdot \n_0  -T^{1-d/2}\calJ\bigg ) 
\bigg \}.
\label{Seff}
\end{eqnarray}

Setting the first variation $\delta S=0$ yields the following Euler-Lagrange equations (see appendix A):
\numparts
\begin{eqnarray}
\label{HD1}
\fl &\frac{\partial q}{\partial t}   =-{\bm \nabla} \cdot \bigg [ 2Dq {\bm \nabla}p  -D{\bm \nabla}q \bigg ],\quad \x\in \Omega\backslash \calU,\\
\label{HD2}
\fl &\frac{\partial p}{\partial t} 
=-D\bigg [{\bm \nabla}^2p  +{\bm \nabla}p \cdot {\bm \nabla }p   \bigg ],\quad \x\in \Omega\backslash \calU,\\
\label{HD3}
\fl& {\bm   j}(\y,t)\cdot \n_0(\y)=\kappa  q(\y,t) ,\quad D{\bm \nabla} p(\y,t)\cdot \n_0(\y)=-\kappa\bigg (p(\y,t)-\lambda\bigg ) , \  \y\in \partial \calU,\\
\fl &q(\y,t)=\owp,\quad p(\y,t)=0,\quad \y \in \partial \Omega,
\label{HD4}
\end{eqnarray}
\endnumparts
with the flux along the optimal path given by
\begin{equation}
\label{HD5}
{\bm j}(\x,t)= 2D q(\x,t){\bm \nabla} p(\x,t)   -D{\bm \nabla} q(\x,t) .
\end{equation}
\endnumparts
These are supplemented by 
 the initial/final conditions
 \begin{equation}
  \label{HD6}
 q(\x,0)=\rho_0 \delta(\x-\x_0)\Theta(h),\quad p(\x,1)=0,
 \end{equation}
and the constraint (\ref{conJ}).
Note that equations (\ref{HD1}) and (\ref{HD2}) can be rewritten as the Hamiltonian dynamical system
\begin{equation}
\frac{\partial q }{\partial t}=\frac{\delta H}{\delta p },\quad\frac{\partial p }{\partial t}=-\frac{\delta H}{\delta q },
\end{equation}
with the Hamiltonian given by
\begin{equation}
H[q,p]=D\int_{\Omega} \bigg [q{\bm \nabla}p\cdot {\bm \nabla}p- {\bm \nabla}q\cdot {\bm \nabla}p \bigg ]d\x .
\end{equation}
In addition, along the optimal path, integration by parts combined with equations (\ref{HD1}), (\ref{HD4}) and the constraint (\ref{conJ}) yields
\begin{eqnarray}
S &\simeq D\int_0^1 dt  \int_{\Omega\backslash \calU} dx\,q(\x,t) {\bm \nabla} p(\x,t)\cdot  {\bm \nabla} p(\x,t).
\label{oct}
\end{eqnarray}

\subsection{Linearised MFT equations and current fluctuations} In the limit $\kappa\rightarrow \infty $, the target becomes totally absorbing with $q(\y,t)=0$ for all $\y\in \partial \calU$. Equations (\ref{HD1})--(\ref{HD4}) then reduce to the single-target version of the variational system analysed in Ref. \cite{Agranov17} for $\Omega =\R^d$. These authors showed how typical current fluctuations can be approximated by a Gaussian distribution whose variance is determined by linearising the variational equations about the NESS of the corresponding deterministic or averaged system and considering stationary solutions in the large-$T$ limit. (The latter is valid provided the underlying systems satisfies an additivity principle \cite{Derrida09}.) We will assume an analogous result holds for a partially absorbing target surface $\partial \calU$ with $\calU\subset \Omega$ and $0<\kappa <\infty$. 
The NESS for a partially absorbing target is obtained by setting  $p(\x,t)\equiv 0$ in equations (\ref{HD1})--(\ref{HD4}), which implies that $\lambda=0$, and solving the resulting steady-state diffusion equation
\numparts 
\begin{eqnarray}
\label{qstar1}
	&{\bm \nabla}^2 q^*(\x)=0,\  \x\in \Omega \backslash   \calU,\\
	\label{qstar2}
	& D{\bm \nabla}q^*(\y) \cdot \n_0(\y)=-\kappa q^*(\y) ,\quad  \y \in \partial \calU,\\
	\label{qstar3}
	&q^*(\x) =\owp,\quad \x\in \partial \Omega .	\end{eqnarray}
	\endnumparts 
Linearising the steady-state version of equations (\ref{HD1})--(\ref{HD4}) about the NESS by introducing the small perturbations
 \begin{equation}
 \label{pert}
\fl Q(\x)=q(\x)-q^*(\x),\quad P(\x)=p(\x) -p^*(\x),\quad {\bm J}(\x)={\bm j}(\x)-{\bm j}^*(\x),
  \end{equation}
 then yields an additional pair of coupled steady-state diffusion equations of the form
 \numparts
\begin{eqnarray}
 \label{NESSlin1}
\fl &   {\bm \nabla}^2P(\x) =0 , \ \x\in \Omega \backslash    \calU,  \\
\label{NESSlin2}
 \fl  &  D{\bm \nabla }P(\y)\cdot \n_0(\y)=-\kappa(P(\y)-\lambda) \quad  \y\in\partial \calU,\quad P(\y)=0,\quad \y \in \partial \Omega ,\\
\label{NESSlin3}
\fl  &{\bm \nabla}^2Q(\x)=2 {\bm \nabla}q^*(\x)\cdot {\bm \nabla}P(\x), \quad \x\in \Omega \backslash    \calU , \\
  \label{NESSlin4}
 \fl &{\bm J}(\y)\cdot \n_0(\y)\equiv - D\bigg ({\bm \nabla}Q(\y)-2q^*(\y){\bm \nabla }P(\y) \bigg )\cdot \n_0(\y)=\kappa Q(\y),\ \y \in \partial \calU,\\
\label{NESSlin5}
\fl  &  Q(\y)=0 ,\ \y \in \partial \Omega.
\end{eqnarray}
\endnumparts

We now show how the linearised MFT equations can be used to characterise typical current fluctuations at the partially absorbing target. First, imposing the steady-state version of condition (\ref{conJ}),
\begin{equation}
\int_{\partial \calU}d\y\, [{\bm j}^*(\y)+{\bm J}(\y)]\cdot \n_0(\y)   \simeq T^{1-d/2}\calJ,
\end{equation}
and using the Robin boundary condition, we have
 \begin{equation}
 \label{2conNh}
\kappa \int_{\partial \calU }Q(\y)d\y = T^{1-d/2}\delta \calJ, \end{equation}
where $\delta \calJ=\calJ-\calJ^*$ and $\calJ^*$ is the deterministic current.
Second, substituting the perturbation expansions (\ref{pert}) into the action (\ref{oct}),
and rewriting the latter in terms of surface integrals gives
  \begin{eqnarray}
S &\simeq  D  \int_{\Omega\backslash \calU} d\x\, q^*(\x) [{\bm \nabla} P(\x)]^2\nonumber \\
&=\frac{1}{2}  \int_{\Omega\backslash \calU} d\x\, {\bm \nabla }P(\x)\cdot \bigg [{\bm J}(\x)+D{\bm \nabla}Q(\x)\bigg ]\nonumber \\
&=\frac{1}{2}  \int_{\Omega\backslash \calU} d\x\, \bigg [{\bm \nabla }\cdot [P(\x){\bm J}(\x)+DQ(\x) {\bm \nabla }P(\x)\bigg ]\nonumber \\
&=\frac{1}{2}  \int_{\partial \calU} d\y\,\bigg  [P(\y){\bm J}(\y)+DQ(\y) {\bm \nabla }P(\y)\bigg ]\cdot \n_0(\y).
\nonumber \\
&=\frac{1}{2}   \int_{\partial \calU} d\y\, Q(\y)\bigg  [\kappa P(\y) +D {\bm \nabla }P(\y)\cdot \n_0(\y)\bigg ]\nonumber \\
&=\frac{\kappa\lambda }{2}  \int_{\partial \calU} d\y\, Q(\y)=\frac{ \lambda T^{1-d/2}}{2}\delta \calJ .
\label{S1target}
\end{eqnarray}
There are no contributions from surface integrals over $\partial \Omega$ since $P(\y)=0=Q(\y)$ for all $\y \in \partial \Omega$. The remaining steps involve deriving a relationship between the Lagrange multiplier $\lambda$ and $\delta \calJ$ by solving the Robin BVPs for $q^*(\x)$, $P(\x)$ and $Q(\x)$ using the spectral decomposition of an associated Dirichlet-to-Neumann (D-to-N) operator acting on the function space $L^2(\partial \calU)$. This is a well established method for dealing with Robin boundary conditions \cite{Grebenkov20d}.

We begin by setting
\begin{eqnarray}
 \label{qxh}
q^*(\x)=\owp(1-\phi(\x) ),
\end{eqnarray}
so that equations (\ref{qstar1})--(\ref{qstar3}) become
\numparts
\begin{eqnarray}
\label{phistar1}
	&{\bm \nabla}^2 \phi(\x)=0,\  \x\in \Omega \backslash   \calU,\\
	\label{phistar2}
	& D{\bm \nabla}\phi(\y) \cdot \n_0(\y)=-\kappa [\phi(\y)-1] ,\quad  \y \in \partial \calU,\\
	\label{phistar3}
	&\phi(\x) =0,\quad \x\in \partial \Omega .	\end{eqnarray}
	\endnumparts
Let $G(\x,\y)$ denote the Dirichlet Green's function for the Laplacian on $\Omega\backslash \calU$, which is defined by
\begin{eqnarray}
\fl &D{\bm \nabla}^2 G(\x,\x')=-\delta(\x-\x'),\quad \x,\x' \in \Omega,\quad G(\x,\x')  =0,  \quad \x\in \partial \Omega \cup \calU
\label{G0}
\end{eqnarray} 
for fixed $\x'$. 
Replacing the boundary condition (\ref{phistar2}) by the Dirichlet condition $\phi(\y)=\Phi(\y)$, $\y \in \partial \calU$, we obtain the formal solution
 \begin{eqnarray}
 \label{nab}
\phi(\x)=-D\int_{\partial \calU}d\z\, \Phi(\z){\bm \nabla}_{\z}G(\z,\x)\cdot \n_0(\z) .
\end{eqnarray}
Imposing the boundary condition (\ref{phistar2}) then yields the following integro-differential equation for $\Phi$:
 \begin{equation}
\label{Phi0}
 {\mathbb L}_0[\Phi](\y)+\frac{\kappa}{D}[\Phi(\y)-1]=0,\quad \y \in \partial \calU,
\end{equation}
where ${\mathbb L}_0$ is the so-called D-to-N operator acting on the space $L^2(\partial \calU)$ \cite{Grebenkov20d}:
\begin{eqnarray}
\label{DtoN2}
  {\mathbb L}_0[f](\y) &=-D{\bm \nabla}_{\y}\bigg [\int_{\partial \calU}d\z\, {\bm \nabla}_{\z} G(\z,\y)\cdot \n_0(\z) f(\z)\bigg ]\cdot \n_0(\y) .
\end{eqnarray}
For a compact surface $\partial \calU$ there exists a countable sets of positive eigenvalues $\mu_n$ and eigenfunctions $v_n(\x)$ satisfying
\begin{equation}
\label{eigL0}
{\mathbb L}_0 v_n(\y)=\mu_nv_n(\y),\quad \y \in \partial \calU.
\end{equation}
Equation (\ref{Phi0}) can now be solved by substituting for $\Phi$ using the eigenfunction expansion
\begin{equation}
\label{Phi2}
\Phi(\y)=\sum_{m=0}^{\infty}\Phi_{m } v_m(\y)
\end{equation}
and taking the inner product with $v_n(\y)$. (We use a real representation of the eigenfunctions such that $\int_{\partial \calU}d\y\, v_n(\y)\, v_m(\y)=\delta_{n.m}$.)
This yields 
\begin{equation}
 \Phi_{n}=\frac{\kappa/D}{\mu_n+\kappa/D} \overline{v}_n,\quad \overline{v}_n=\int_{\partial \calU}v_n(\y)d\y.
\end{equation}
Combining with equation (\ref{nab}) shows that
\begin{equation}
\phi(\x)=\sum_{n}\frac{\kappa/D}{\mu_n+\kappa/D}\overline{v}_n\phi_n(\x),
\label{pixie1}
\end{equation}
and
\begin{equation}
 \phi_n(\x)=-D\int_{\partial \calU} d\y\, v_n(\y) {\bm \nabla} _{\y}G(\y,\x)\cdot \n_0(\y).
 \label{pin}
\end{equation}
On the target surface,
\begin{equation}
q^*(\y)=\owp(1-\Phi(y)),\quad \Phi(\y) = \sum_{n}\frac{\kappa/D}{\mu_n+\kappa/D}\overline{v}_nv_n(\y).
\label{pixie2}
\end{equation}
Finally, comparing equations (\ref{NESSlin1}) and (\ref{NESSlin2})  with (\ref{phistar1})--(\ref{phistar3}), it follows that
 \begin{eqnarray}
 \label{Pxh}
P(\x)=\lambda \phi(\x).
\end{eqnarray}

Next we replace the boundary condition (\ref{NESSlin4}) by the Dirichlet boundary condition
$Q(\y)=\Psi(\y)$ for $ \y \in \partial \calU$
and write down the formal solution of equation (\ref{NESSlin3}):
\begin{eqnarray}
  Q(\x)&= -2D  \int_{\Omega\backslash \calU}d\z\, G(\x,\z){\bm \nabla}\cdot \bigg [q^*(\z) {\bm \nabla}P(\z)\bigg ] \nonumber \\
  &\quad -D\int_{\partial \calU}d\z\,  \Psi(\z){\bm \nabla}_{\z} G(\z,\x)\cdot \n_0(\z) .
\end{eqnarray}
We have used the identity 
\[{\bm \nabla}\cdot \bigg [q^*{\bm \nabla}P\bigg ]={\bm \nabla} q^*\cdot {\bm \nabla P}+q^*{\bm \nabla}^2 P\]
with ${\bm \nabla}^2 P=0$.
Imposing the boundary condition (\ref{NESSlin4}) then yields the following integro-differential equation for $\Psi$:
\begin{eqnarray}
\label{Lin0}
\fl   {\mathbb L}_0[\Psi](\y)+\frac{\kappa}{D} \Psi(\y) &=-2D   \int_{\Omega\backslash \calU}d\z\, \bigg ( {\bm \nabla}_{\y}G(\y,\z)\cdot \n_0(\y)\bigg ){\bm \nabla}\cdot \bigg [q^*(\z) {\bm \nabla}P(\z)\bigg ]\nonumber \\
\fl &\quad +2q^*(\y){\bm \nabla }P(\y)  \cdot \n_0(\y) ,\,\  y \in \partial \calU.
\end{eqnarray}
Introducing the eigenvalue expansion
\begin{equation}
\Psi(\y)=\sum_{n=0}^{\infty}\Psi_nv_n(\y),
\end{equation}
multiplying both sides of equation (\ref{Lin0}) by $v_n(\y)$ and integrating with respect to $\y \in \calU$, we have 
\begin{eqnarray}
\bigg (\mu_n+\frac{\kappa}{D}\bigg )\Psi_{n}&=2 \int_{\Omega\backslash \calU}d\z\, \phi_n(\z){\bm \nabla}\cdot \bigg [q^*(\z) {\bm \nabla}P(\z)\bigg ]\nonumber \\
  &\quad  +2\int_{\partial \calU}d\y\, v_n(\y)q^*(\y){\bm \nabla }P(\y)  \cdot \n_0(\y).
\end{eqnarray}
The presence of $\phi_n(\z)$ on the right-hand side follows from equation (\ref{pin}).
Moreover, equation (\ref{2conNh}) implies that
\begin{eqnarray}
  \fl  &T^{1-d/2}\delta \calJ \nonumber \\
 \fl &\equiv  \kappa \int_{\partial \calU}\Psi(\y)d\y=\kappa\sum_{n}\Psi_n \overline{v}_n \nonumber \\
 \fl  &=-2D \int_{\Omega\backslash \calU}d\z\, \phi(\z){\bm \nabla}\cdot \bigg [q^*(\z) {\bm \nabla}P(\z)\bigg ] +2D\int_{\partial \calU}d\y\, \Phi(\y)q^*(\y){\bm \nabla }P(\y)  \cdot \n_0(\y)\nonumber \\
 \fl  & =-2D \int_{\Omega\backslash \calU}d\z\, \phi(\z){\bm \nabla}\cdot \bigg [q^*(\z) {\bm \nabla}P(\z)\bigg ] +2D\int_{\Omega \backslash \calU}d\y\,{\bm \nabla}\cdot  \bigg [\phi(\z)q^*(\y){\bm \nabla }P(\y) \bigg ]\nonumber \\
 \fl&=2D\int_{\Omega \backslash \calU}d\x\, q^*(\x){\bm \nabla}\phi(\x) \cdot {\bm \nabla }P(\x)  
 =2D\lambda \int_{\Omega \backslash \calU}d\x\, q^*(\x){\bm \nabla}\phi(\x) \cdot {\bm \nabla }\phi(\x) . 
 \label{nabka}
\end{eqnarray}
The third line follows from equations (\ref{Phi2})--(\ref{pin}), the divergence theorem has been used on the fourth line, and we have set $P(\x)=\lambda \phi(\x)$ on the final line. Interestingly, equation (\ref{nabka}) is identical in form to the analogous result derived in Ref. \cite{Agranov17} for ta totally absorbing target.

Finally, combining equations (\ref{LDP0}), (\ref{S1target}) and (\ref{nabka}) leads to equation (\ref{LDP}), after  converting back to unscaled coordinates, with
\begin{equation}
\calC(\kappa)=2D\int_{\Omega \backslash \calU}d\x\, q(\x){\bm \nabla}\phi(\x) \cdot {\bm \nabla }\phi(\x)  .
\label{Ckap}
\end{equation}
The typical size of the total fluctuations in the large-$T$ limit is then characterised by the $\kappa$-dependent variance $\calC(\kappa )/T$. 
However, in order to calculate $\calC(\kappa)$ for a general geometric configuration, we still have to determine the spectrum of the D-to-N operator ${\mathbb L}_0$ and evaluate the integral on the right-hand side of equation (\ref{Ckap}) for $q^*(\x)=\owp(1-\phi(\x))$  and $\phi(\x)$ given by equation (\ref{pixie1}). We now consider two examples where explicit solutions can be obtained without recourse to spectral theory, namely, a finite interval and a circular annulus, respectively. 

\subsection{Current fluctuations in an interval} 

\begin{figure}[b!]
\centering
\includegraphics[width=10cm]{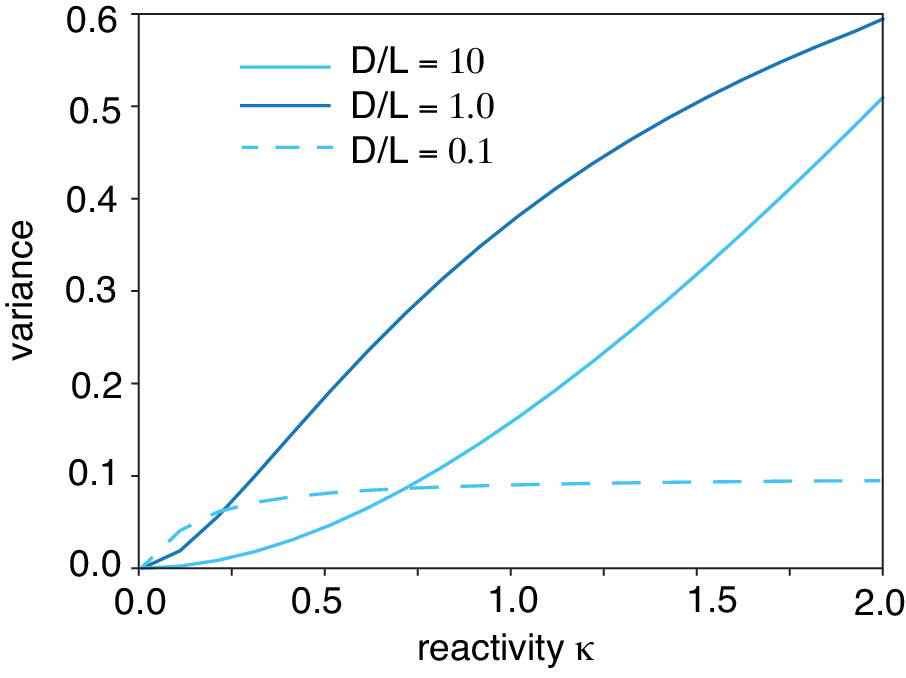} 
\caption{Plots of $\mbox{Var}[\calJ(T)]$ (in units of $\owp/T$) against the reactivity $\kappa_0$ for various interval lengths $L$, where $\calJ(T)$ is the current into a target at $x=0$.}
\label{fig2}
\end{figure}

Consider a non-interacting Brownian gas in the finite interval $[0,L]$ with a partially absorbing boundary at $x=0$ and a fixed density $\wp=\owp$ at the end $x=L$, see Fig. \ref{fig1}(a). 
Equation (\ref{Ckap}) reduces to 
\begin{equation}
\calC(\kappa)=2\owp D\int_0^Ldx\, (1-\phi(x)) \left (\frac{d\phi(x)}{dx}\right )^2 ,
\label{1DCkap}
\end{equation}
with $\phi(x)$ satisfying the BVP (after noting that $\n_0\cdot {\bm \nabla}\phi(\y)  \rightarrow -\phi'(0)$)
\begin{eqnarray} 
 \label{noshh}
 \frac{d^2  \phi(x)}{dx^2}=0,\quad D\left . \frac{d\phi(x)}{d x}\right |_{x=0}=\kappa  (\phi(0)-1),\quad \phi(L)=0 .
 \end{eqnarray}
Equation (\ref{noshh}) has the solution
\begin{equation}
 \phi(x)=  \frac{ L-x}{L}\frac{\kappa/D}{\kappa/D+1/L}.
 \end{equation}
 Comparison with the general result (\ref{pixie1}) shows that in the 1D case the D-to-N operator is a scalar with a single eigenvalue $\mu =1/L$. It will be convenient to rewrite $\phi(x)$ in the more compact form
 \begin{equation}
 \phi(x)=\frac{ L-x}{L}\frac{1}{1+\gamma},\quad \gamma =\frac{D}{\kappa L}.
 \end{equation}
Plugging the solution for $\phi(x)$ into equation (\ref{1DCkap}) then gives
 \begin{eqnarray}
\calC(\kappa)&=\frac{\owp D}{L^2(1+\gamma)^3}\int_0^Ldx\, \bigg (\gamma+\frac{  x}{L} \bigg )
  =\frac{\owp D(1+2\gamma)}{L(1+\gamma)^3} \nonumber \\
  &=\frac{ \owp D}{L}   \frac{ 1+2D/(\kappa L)}{(1+D/(\kappa L))^3}.
\end{eqnarray}
In Fig. \ref{fig2} we plot the variance $\mbox{Var}[\calJ(T)]=\langle (\calJ(T)-\calJ^*)^2\rangle=\calC(\kappa)/T$
as a function of the reactivity $\kappa$ and various values of $D/L$. (The corresponding deterministic flux is $\calJ^*=\kappa\owp (D/L)/(\kappa+(D/L))$.) It can be seen that the variance is a monotonically increasing function of the reactivity $\kappa$ that asymptotes in the totally absorbing limit to
\begin{equation}
\lim_{\kappa\rightarrow \infty}\calC(\kappa)=\frac{\owp D}{L} .
\end{equation}

\subsection{Current fluctuations in a circular annulus}

\begin{figure}[b!]
\centering
\includegraphics[width=10cm]{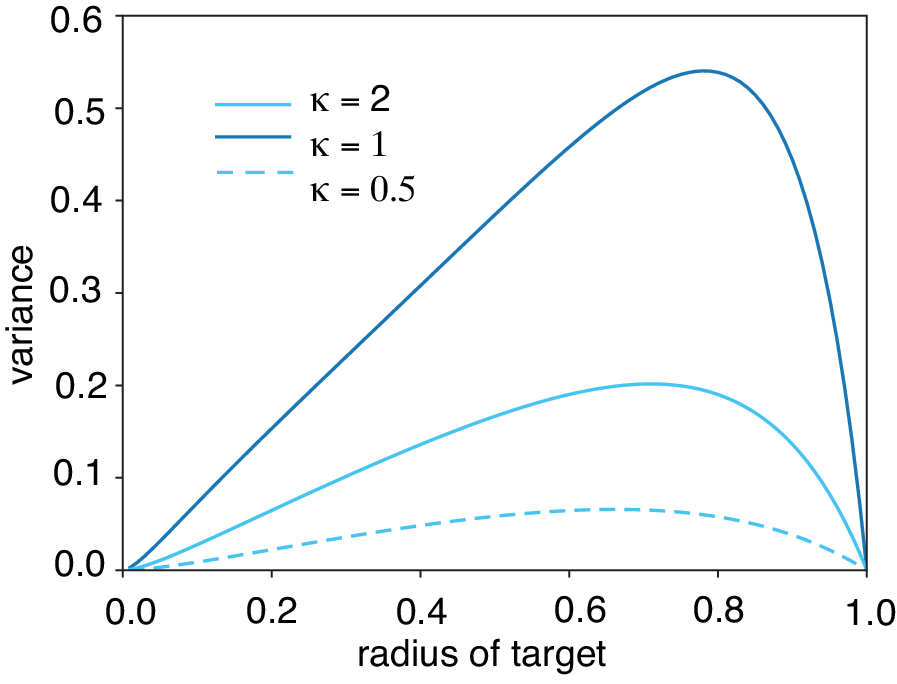} 
\caption{Plot of variance $\mbox{Var}[\calJ(T)]$ (in units of $2\pi \owp/T$) against the radius $R$ for various reactivities $\kappa_0$, where $\calJ(T)$ is the current into a circular target of radius $R$ centred in the unit disc.}
\label{fig3}
\end{figure}

Now suppose that $\partial \calU$ and $\partial \Omega$ are concentric circles of radii $R$ and $L$, respectively, with $R<L$. Introducing the polar coordinates $(r,\theta)$ and exploiting circular symmetry, equation (\ref{Ckap}) becomes
\begin{equation}
\calC(\kappa)=4\pi \owp D\int_R^Ldr\, r(1-\phi(r)) \left (\frac{d\phi(r)}{dr}\right )^2 ,
\label{2DCkap}
\end{equation}
with
\numparts 
\begin{eqnarray}
\label{qpol1}
	&\frac{1}{r}\frac{d}{dr}r\frac{d\phi(r)}{dr} =0,\  R<r<L,\\
	\label{qpol2}
	&D\left . \frac{d\phi(r)}{dr}\right |_{r=R}=\kappa[ \phi(R) -1],\quad \phi(L)=0.\end{eqnarray}
	\endnumparts 
The solution is
\begin{equation}
\label{qsol}
\phi(r)=\Gamma_0\ln(r/L),\quad \Gamma_0=-\frac{1}{D/(\kappa R)+\ln(L/R)}.
\end{equation}
 Plugging the solution for $\phi(r)$ into equation (\ref{2DCkap}) then gives
\begin{eqnarray}
\calC(\kappa)&=4\pi \owp D\int_R^Ldr\, r\bigg (1+\Gamma_0\ln(r/L)\bigg ) \left (\frac{\Gamma_0}{r}\right )^2\nonumber \\
&=4\pi \owp D\Gamma_0^2\bigg [\ln(L/R)-\frac{\Gamma_0}{2}\ln^2L/R\bigg ]\nonumber \\
&=\frac{2\pi   \owp D}{T}  \frac{  {2D}/(\kappa R) +\ln(L/R)}{ [D/(\kappa R)+ \ln(L/R)]^3}  \ln(L/R).
\label{varran}
\end{eqnarray}
 In Fig. \ref{fig3} we plot the variance $C(\kappa)/T$ (in units of $2\pi \owp/T$)
as a function of the target radius $R$ and various values of $\kappa$ with $D=L=1$. Again we find that the variance is a monotonically increasing function of $\kappa$. On the other hand, the variance is a unimodal function of $R$ with a maximum at a value $R^*(\kappa)$, $0 <R^*(\kappa)<1$, that is an increasing function of $\kappa$.

\section{Multiple partially absorbing targets and cross-correlations in the current statistics}

\begin{figure}[b!]
\centering
\includegraphics[width=10cm]{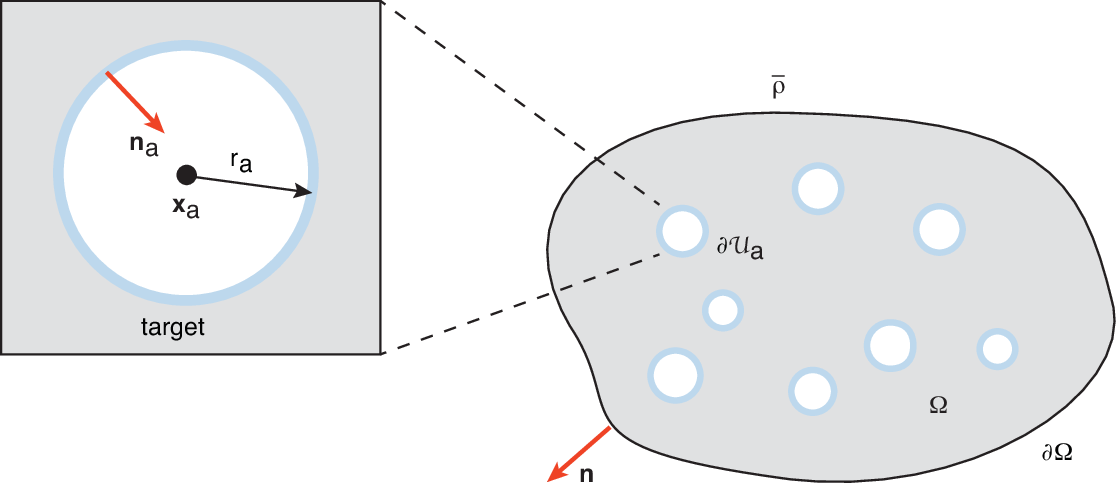} 
\caption{Particles diffusing in a 2D bounded domain $\Omega$ containing $M$ circular targets $\calU_a$ , $a=1,\ldots,M$. The $a$-th target is centred at $\x_a$, with a radius $r_a$ and a unit normal $\n_a$.}
\label{fig4}
\end{figure}

Now suppose that the bounded domain $\Omega \subset \R^d$, $d>1$, contains a set of $M$ partially absorbing targets $\calU_a$, $a=1,\ldots,M$. (The example of circularly symmetric targets in $\Omega \subset \R^2$ is shown in Fig. \ref{fig4}.) This general setup differs in a number of respects from the multi-target problem analysed in Ref. \cite{Agranov17}. In particular, the latter authors considered a set of totally absorbing patches on an otherwise reflecting surface $\partial \calU$ with $\calU \subset \R^3$. One of the main results of Ref. \cite{Agranov17} was a general expression for the covariance matrix of multi-target current fluctuations, which was obtained using potential theory. This then established that cross-correlations in the fluctuating current statistics only occur in the presence of particle-particle interactions, irrespective of the geometry of the underlying system. Extending the analysis of Ref. \cite{Agranov17}, we will show that cross-correlations can occur when the targets are partially absorbing. We then illustrate the theory in section 5 by calculating the covariance matrix in the small target limit using singular perturbation theory. For simplicity, we take the reactivity to be the same at each target surface. 

\newpage

\subsection{Variational formulation}

Consider the DK equations (\ref{RBDK1}) and (\ref{RBDK2}) in the case of the configuration shown in Fig. \ref{fig4} with $\calU=\cup_{a=1}^M \calU_a$. Let $\calN_{a}(T)$ denote the stochastic number of particles absorbed by the $a$-th target over the time interval $[0,T]$ and set
\begin{equation}
\label{nhsmulti}
\calJ_a(T)=\frac{\calN_{a}(T)}{T}=\frac{1}{T}\int_0^T \bigg [\int_{\partial \calU_a} {\bm \frJ}(\y,t)\cdot \n_a(\y)d\y\bigg ]dt,
\end{equation}
where $\n_a$ is the inward unit normal of the $a$-th target. We wish to determine the statistics of the multiple target fluctuations in the large-$T$ limit.
The variational construction proceeds along identical lines to section 3, except that there are now $M$ Lagrange multipliers $\lambda_a$ implementing the $M$ constraints (after rescaling)
\begin{equation}
\label{conNhmulti}
\int_0^1 \bigg [\int_{\partial \calU_a}{\bm j}(\y,t)\cdot \n_a(\y) d\y \bigg ]dt =T^{1-d/2}\calJ_a 
\end{equation}
for $a =1,\ldots M$. The flux ${\bm j}$ is given by equation (\ref{HD3}). The resulting path integral representation of the sum over histories yields the following multi-target version of equations (\ref{HD1})--(\ref{HD4}):
\numparts
\begin{eqnarray}
\label{multi1}
\fl &\frac{\partial q}{\partial t}   =-{\bm \nabla} \cdot {\bm j} ,\\
\label{multi2}
\fl &\frac{\partial p}{\partial t} 
=-D\bigg [{\bm \nabla}^2p  +[ {\bm \nabla}p]^2   \bigg ],\quad \x\in \Omega\backslash \cup_{a=1}^M \calU_a,\\
 \label{multi3}
\fl & {\bm j}(\y,t)\cdot \n_a(\y)=\kappa   q(\y,t),\ {\bm \nabla }p(\y,t)\cdot \n_a(\y)=-\kappa  \bigg (p(\y,t)-\lambda_a\bigg ) , \ \y\in \partial \calU_a,\\
\fl &q(\y,t)=\owp,\quad p(\y,t)=0,\quad \y \in \partial \Omega.
\label{multi4}
\end{eqnarray}
\endnumparts

Setting $p\equiv 0$ and time derivatives to zero in equations (\ref{multi1})--(\ref{multi4}) gives the steady-state diffusion equation for the multi-target NESS $q^*$:
\numparts 
\begin{eqnarray}
\label{mstar1}
	&{\bm \nabla}^2 q^*(\x)=0,\  \x\in \Omega \backslash \cup_{a=1}^N  \calU_a,\\
	\label{mstar2}
	&D{\bm \nabla}q^*(\y) \cdot \n_a(\y)=-\kappa q^*(\y) ,\quad  \y\in\partial \calU_a,\quad a=1,\ldots,M\\
	&q^*(\y) =\owp,\quad \y\in \partial \Omega. 	
	\label{mstar3}\end{eqnarray}
	\endnumparts 
Similarly, introducing the small perturbations (\ref{pert}), we obtain the multi-target version of equations (\ref{NESSlin1})--(\ref{NESSlin5}):
 \numparts
\begin{eqnarray}
\label{HDlin1}
\fl &   {\bm \nabla}^2P(\x) =0 , \quad \x\in \Omega \backslash \cup_{a=1}^N  \calU_a,  \\
\label{HDlin2}
\fl  &D{\bm \nabla }P(\y)\cdot \n_a(\y)=-\kappa  \bigg (P(\y)-\lambda_a\bigg ) ,\quad  \y\in\partial \calU_a,\quad P(\y)=0,\quad \y \in \partial \Omega, \\
\label{HDlin3}
\fl  &{\bm \nabla}^2Q(\x)=2{\bm \nabla}q^*(\x)\cdot {\bm \nabla}P(\x), \quad \x\in \Omega \backslash \cup_{a=1}^N  \calU_a,\\
\fl   &{\bm J}(\y)\cdot \n_a(\y)\equiv D\bigg [2 q^*(\y){\bm \nabla}P(\y)- {\bm \nabla}Q(\y) \bigg ]\cdot \n_a(\y)=\kappa  Q(\y),\quad \y \in \partial \calU_a,
\label{HDlin4}\\
\fl & Q(\y)=0,\quad \y \in \partial \Omega.
\label{HDlin5}
\end{eqnarray}
\endnumparts
Also note that imposing the steady-state version of the condition (\ref{conNhmulti}),
 \begin{equation}
\label{2conNhmulti}
\int_{\partial \calU_a}[{\bm j}^*(\y)+{\bm J}(\y)]\cdot \n_a(\y) d\y =T^{1-d/2}\calJ_a,
\end{equation}
implies that
 \begin{equation}
 \label{Fmulti}
\int_{\partial \calU_a} d\y\, {\bm J}(\y)\cdot \n_a(\y)=\kappa  \int_{\partial \calU_a}   Q(\y) d\y  =T^{1-d/2}\delta \calJ_a, 
\end{equation}
where $ \calJ_a =\calJ_a-\calJ_a^*$.

\subsection{Potential theory} Similar to Ref. \cite{Agranov17}, we introduce the ``potential'''$\phi_{\ell}$ that satisfies Laplace's equation
\numparts 
\begin{eqnarray}
\label{pot1}
	&{\bm \nabla}^2 \phi_{\ell}(\x)=0,\  \x\in \Omega \backslash \cup_{a=1}^N  \calU_a,\\
	\label{pot2}
	&D{\bm \nabla}\phi_{\ell}(\y) \cdot \n_a(\y)=-\kappa [ \phi_{\ell}(\y)-\delta_{a,\ell}] ,\quad  \y\in\partial \calU_{a},\\
	& \phi_{\ell}(\y) =0,\quad \y\in \partial \Omega. 	
	\label{pot3}\end{eqnarray}
	\endnumparts 
From the principle of superposition, we have
\begin{equation}
q^*(\x)=\owp \bigg [1-\sum_{\ell=1}^M\phi_{\ell}(\x)\bigg ],\quad P(\x)= \sum_{\ell=1}^M\lambda_{\ell}\phi_{\ell}(\x).
\label{qP}
\end{equation}
Substituting these expressions into the action (\ref{oct}) gives 
  \begin{eqnarray}
  \label{Smtarget}
S &\simeq  \frac{1}{2}\sum_{\ell,\ell'=1}^M\lambda_{\ell}\lambda_{\ell'} \Sigma_{\ell\ell'},
\end{eqnarray}
where ${\bm \Sigma}$ is a symmetric $M\times M$ matrix with
\begin{equation}
\Sigma_{\ell\ell'}=2D\int_{\Omega\backslash \calU} d\x\, q^*(\x) {\bm \nabla} \phi_{\ell}(\x) \cdot {\bm \nabla}\phi_{\ell'} (\x) .
\label{thet}
\end{equation}
Following along similar lines to appendix D of Ref. \cite{Agranov17}, we see that
\begin{eqnarray}
\fl  \Sigma_{\ell\ell'}&=D  \int_{\Omega\backslash \calU} d\x\,\phi_{\ell}(\x) \phi_{\ell'}(\x)  {\bm \nabla}^2q^*(\x) \nonumber \\
\fl  & \quad +D \int_{\Omega\backslash \calU} d\x\,  {\bm \nabla }\cdot \bigg (q^*(\x){\bm \nabla}[\phi_{\ell}(\x)\phi_{\ell'}(\x)]-\phi_{\ell}(\x)\phi_{\ell'}(\x){\bm \nabla}q^*(\x)\bigg )\nonumber \\
\fl  &=D \int_{\partial \calU\cup \partial \Omega} d\y\,\bigg (q^*(\y){\bm \nabla}[\phi_{\ell}(\y)\phi_{\ell'}(\x)]-\phi_{\ell}(\y)\phi_{\ell'}(\y){\bm \nabla}q^*(\y)\bigg )\cdot \n_0(\y). 
\label{Smtarget0}
\end{eqnarray}
We have used the divergence theorem and the fact that $q^*(\x)$ and $\phi_{\ell}(\x)$ are harmonic functions. The contributions from the surface integrals on $\partial \Omega$ vanish since $\phi_{\ell}(\y)=0$ for all $\y \in \partial \Omega$. Hence,
\begin{eqnarray}
   \Sigma_{\ell\ell'} &=D\sum_{a=1}^M   \int_{\partial \calU_a} d\y\,\bigg (q^*(\y)\bigg [\phi_{\ell}(\y){\bm \nabla}\phi_{\ell'}(\y)+\phi_{\ell'}(\y){\bm \nabla}\phi_{\ell}(\y)\bigg ]\nonumber \\
 &\hspace{3cm}-\phi_{\ell}(\y)\phi_{\ell'}(\y){\bm \nabla}q^*(\y)\bigg )\cdot \n_0(\y).
 \label{wow}
\end{eqnarray}
Clearly the off-diagonal terms ($\ell\neq \ell'$) vanish when the targets are totally absorbing since $q^*(\y)=0$ for all $\y\in \partial \calU_a$, $a=1,\ldots,M$, and $\phi_{\ell}(\y)=0$ for all $\y \in \partial \calU_a$ and $a\neq \ell$. That is,
\begin{eqnarray}
   &  \Sigma_{\ell\ell'}= -   D\delta_{\ell,\ell'} \int_{\partial \calU_{\ell}} d\y\ \phi^2_{\ell}(\y){\bm \nabla}q^*(\y) \cdot \n_0(\y)  .
\label{wow2}
\end{eqnarray}
This recovers the general result of \cite{Agranov17}. On the other hand, the off-diagonal terms are generally non-zero when the targets are partially absorbing.

Extending the spectral analysis of section 3.2 to multiple targets (see appendix B), it can be shown that the Lagrange multipliers $\{\lambda_a\}$ and current fluctuations $\{\delta \calJ_a\}$ are related according to 
\begin{eqnarray}
\fl T^{1-d/2}\delta \calJ_a=2D\sum_{\ell=1}^M \lambda_{\ell} \int_{\Omega \backslash \calU}d\x\, q^*(\x){\bm \nabla}\phi_a(\x) \cdot {\bm \nabla }\phi_{\ell}(\x)\equiv \sum_{\ell=1}^M\Sigma_{a\ell} \lambda_{\ell}.
\label{nabka2}
\end{eqnarray}
Combining equations (\ref{LDP0}), (\ref{Smtarget}), (\ref{nabka2}) then leads to the multi-target distribution (\ref{mLDP}) with ${\bm \Sigma}/T$ equal to the covariance matrix, that is,
\begin{equation}
\langle \delta \calJ_a\delta \calJ_b\rangle =\frac{\Sigma_{ab}}{T}.
\end{equation}
Equation (\ref{wow})establishes one of our major results, namely, there exist off-diagonal terms of the covariance matrix when the targets are partially reflecting. Consequently, there exist corresponding cross-correlations in the current fluctuations of the non-interacting Brownian gas. These cross-correlations vanish in the totally absorbing limit $\kappa \rightarrow \infty$, consistent with Ref. \cite{Agranov17}.

\section{Matched asymptotic analysis of current fluctuations in the small target limit}

In this  section we use singular perturbation theory to analyse current fluctuations for the 2D multi-target configuration shown in Fig. \ref{fig4}, under the additional assumptions that the targets are much smaller than the size of the search domain and that they are well-separated. In particular, the $a$-th target is taken to be a circular disc with centre $\x_a$ and radius $r_a$. That is, $\calU_a=\{\x\in \Omega,\ |\x-\x_a|\leq r_a\}$ for $a=1,\ldots,M$.
We fix the length scale by setting $L:=|\Omega|^{1/2}=1$, and take $r_a=\epsilon R_a$ with $0<\epsilon \ll 1$, $|\x_a-\x_b| =O(1)$ for all $b\neq a$, and $\min_{\y}\{|\x_a-\y|,\y \in \partial \Omega \} =O(1)$ for all $a,b=1,\ldots,M$. As in the single-target case, see section 3, the goal is to solve the steady-state diffusion equations for $q^*(\x)$, $P(\x)$ and $Q(\x)$ in order to determine the relationship between the Lagrange multiplier $\lambda_a$ and current fluctuations at the boundaries $\partial \calU_a$. Applying singular perturbation methods, we derive an asymptotic expansion of each density with respect to the small parameter $\nu=-1/\ln \epsilon$ by matching an inner solution around each target with a corresponding outer solution. We exploit the fact that in the small target regime, we can take the densities to be constant for all $\y \in \calU_a$.

\subsection{Asymptotic analysis of the potential $\phi_{\ell}(\x)$.}

Consider the BVP for the potential $\phi_{\ell}(\x)$ staifying equations (\ref{pot1})--(\ref{pot3}).
In order to construct the inner solution around the $a$-th target, we introduce the stretched polar coordinate $R=|\x-\x_a|/\epsilon$ such that $r_a=\epsilon R_a$. Setting $\phi_l=\Phi_{\ell,a}(R)$ for $\x \sim  \x_a$, we have the inner equation
\numparts
\begin{eqnarray}
\fl &\frac{1}{R}\frac{d}{dR}R\frac{d\Phi_{\ell,a}(R)}{dR}=0,\quad R>R_a ; \quad D \Phi'_{\ell,a}(R_a) =\epsilon \kappa [\Phi_{\ell,a}(R_a) -\delta_{\ell,a} ],
\end{eqnarray}
\endnumparts
 which has the solution
 \begin{equation}
\label{innerP0}
\Phi_{\ell,a}(R)=\nu \calA_{\ell,a} \bigg [ \ln( R /R_a)+\frac{D}{\epsilon \kappa  R_a}\bigg ]+\delta_{\ell,a},\quad R>R_a
\end{equation}
for the unknown constant $\calA_a $\footnote{The factor of $\nu$ in equation (\ref{innerP0}) is introduced so that $\calA_a =O(1)$ with respect to a perturbation expansion in powers of $\nu$. The function in square brackets can be written more compactly as
 $\ln(R/d_a(\kappa))$ where
$d_a(\kappa)=R_a\e^{-D/\kappa R_a}$. From electrostatics, $d_a(\infty)=R_a$ is known as the shape capacitance for a circular target of radius $R_a$ \cite{Ward93}. It is known that a circular domain has the smallest
logarithmic capacitance of all domains with the same area. Our analysis could be extended to non-circular targets by modifying $d_a(\kappa)$ accordingly. 
 }
In order to have a non-trivial dependence on the reactivity in the small-$\epsilon$ limit, we assume that $D=D_0/\nu$ and $\kappa =\kappa_0/\epsilon$ so that
\begin{equation}
\frac{\nu D}{\epsilon \kappa  }= \frac{D_0}{\kappa_0}.
\end{equation}
This ensures that (i) the absorption flux into the target remains non-zero as the target circumference $2\pi \epsilon R_a\rightarrow 0$ and (ii) the deviation from the totally absorbing case occurs at $O(1)$ in the asymptotic expansion. Equation (\ref{innerP0}) becomes
\begin{equation}
\label{innerP}
\fl \Phi_{\ell,a}(R)=\delta_{\ell,a}+\gamma_a \calA_{\ell,a}+\nu \calA_{\ell,a}  \ln( R /R_a),\quad R>R_a,\quad \gamma_a=\gamma_a(\kappa_0)\equiv \frac{D_0}{\kappa_0R_a}.
\end{equation}

The outer solution is obtained by treating $\calU_a$ as a point source/sink, which is equivalent to zooming out over the domain $\Omega$.  The corresponding outer equation takes the form
\numparts
\begin{eqnarray}
&{\bm \nabla}^2 \phi_{\ell}(\x)=0,\ \x\in \Omega\backslash \{\x_1,\ldots,\x_M\}; \quad \phi_{\ell}(\x)=0,\ \x \in \partial \Omega,
\end{eqnarray}
together with the matching condition
\begin{equation}
\label{matchP}
\phi_{\ell}(\x)\sim \delta_{a,\ell}+  \calA_{\ell,a}\bigg [1+\gamma_a+ \nu  \ln (|\x-\x_a|/R_a) \bigg ]
\end{equation}
\endnumparts
as $\x\rightarrow \x_a$. The next step is to introduce the Dirichlet Green's function $G_{\Omega}(\x,\y)$ for the 2D Laplacian on $\Omega \subset \R^2$, which is defined by\footnote{It is important to distinguish $G_{\Omega}(\x,\x')$ from the Dirichlet Green's function $G(\x,\x')$ for the Laplacian on $\Omega\backslash \calU \subset \R^d$, see equations (\ref{G0}).}
\numparts
\label{G1}
\begin{eqnarray}
\fl &D_0{\bm \nabla}^2 G_{\Omega}(\x,\x')=-\delta(\x-\x'),\quad \x,\x' \in \Omega,\quad G_{\Omega}(\x,\x')  =0,  \quad \x\in \partial \Omega
\end{eqnarray}
\endnumparts
for fixed $\x'$. Note that the 2D Green's function $G_{\Omega} $ can be decomposed as
\begin{equation}
G_{\Omega}(\x,\x')=-\frac{ \ln |\x-\x'|}{2\pi D_0}+H_{\Omega}(\x,\x'),
\end{equation}
where $H_{\Omega}$ is the regular (non-singular) part of the Green's function. 
We now make the ansatz 
\begin{equation}
\label{outerP}
\phi_{\ell}(\x)\sim -2\pi  \nu  D_0 \sum_{b=1}^M\calA_{\ell,b}G_{\Omega}(\x,\x_b)
\end{equation}
for $\x \notin \{\x_a,\, a=1,\ldots,M\}$
and match with the inner solution. The outer solution (\ref{outerP}) shows that  as $\x\rightarrow \x_a$,
\begin{eqnarray}
 \fl \phi_{\ell}(\x)&\rightarrow \nu  \calA_{\ell,a}\ln|\x-\x_a|-2\pi D_0\nu  \calA_{\ell,a}  H_{\Omega}(\x_a,\x_a)  
  -2\pi \nu D_0\sum_{b\neq a} \calA_{\ell,b} G_{\Omega}(\x_a,\x_b). 
  \end{eqnarray}
Comparison with the asymptotic limit in equation (\ref{matchP}) implies that
\begin{equation}
\label{AP}
\sum _{b=1}^M \bigg [\delta_{a,b} \bigg (1+\gamma_a\bigg )+  \nu  {\calM}_{ab} \bigg ]\calA_{\ell,b}\sim -\delta_{\ell,a},
\end{equation}
with 
\begin{eqnarray}
\fl  {\calM}_{aa} &=2\pi D_0H_{\Omega} (\x_a,\x_a)- \ln R_a   ,\quad {\calM}_{ab} =2\pi D_0 G_{\Omega}(\x_a,\x_b),\ b\neq a.
\label{calG}
\end{eqnarray}
That is, 
\begin{equation}
\label{calAmatrix}
\calA_{\ell,a}=-[{\bm I}(\kappa_0) +\nu{\bm \calM}]_{a\ell}^{-1},
\end{equation}
and
\begin{equation}
I_{ab}(\kappa_0) =\delta_{a,b} \bigg (1+\frac{D_0}{\kappa_0 R_a}\bigg ).
\end{equation}

It immediately follows from equations (\ref{qP}) that the inner solutions of $q^*(\x)$ and $P(\x)$ around the $a$-th target are
\begin{equation}
q^*_a(R)=\owp \bigg [1-\calB_a \bigg ( \nu \ln( R /R_a)+\gamma_a\bigg ) \bigg ]
\end{equation}
and
\begin{eqnarray}
\fl P_a(R)&=\sum_{a=1}\lambda_a \Phi_{\ell,a}(R) 
=\lambda_a- \wlam_{a} \bigg [ \nu \ln( R /R_a)+\gamma_a \bigg ] ,\quad R>R_a
\end{eqnarray}
with
\begin{equation}
\wlam_a=\sum_{\ell=1}^M[{\bm I}(\kappa_0) +\nu{\bm \calM}]_{a\ell}^{-1}\lambda_{\ell} ,\quad \calB_a=-\sum_{\ell=1}^M[{\bm I}(\kappa_0) +\nu{\bm \calM}]_{a\ell}^{-1}.
\label{Fmatrix}
\end{equation}
Note that the effective Lagrange multiplier $\widehat{\lambda}_a$ is a non-perturbative function of the small parameter $\nu$, which was obtained by matching the inner and outer solutions using Green's functions along the lines originally developed in Refs. \cite{Ward93,Ward93a}. This effectively sums over the logarithmic terms, which is equivalent to calculating the asymptotic solution for all terms of $O(\nu^k)$ for any $k$. (The expression for the coefficient $\calB_a$ is also non-perturbative with respect to $\nu$.)
Moreover, $\widehat{\lambda}_a$ is a non-local function of the set of bare Lagrange multipliers with the coupling mediated by the symmetric Green's function matrix ${\bm \calM}$. Given the inner solutions for $q^*(\x)$ and the potentials $\phi_{\ell}(\x)$ we could calculate the covariance matrix for current fluctuations using (\ref{wow}). We will follow a different approach here by applying singular perturbation theory to calculate $Q(\x)$ directly.

\subsection{Asymptotic analysis of $Q(\x)$}

The inner equation for $Q=Q_{a}(R) $ around the $a$-th target is
\numparts
 \begin{eqnarray}
  &\frac{1}{R}\frac{d}{dR}R\frac{dQ_{a}(R)}{dR}=- \frac{2\nu^2 \owp\widehat{\lambda}_a \calB_a }{R^2} ,\quad R>R_a,\\
  & D_0Q'_{a}(R_a)+ \frac{2\nu D \owp \gamma_a\widehat{\lambda}_a \calB_a }{R_a}=\nu \kappa_0 Q_a(R_a).
\end{eqnarray}
\endnumparts
We have used the corresponding inner solutions for $q^*(\x)$ and $P(\x)$, which imply that
\begin{equation}
{\bm \nabla}P(\x)\cdot \n_0(\x) \rightarrow \frac{\nu \wlam_a}{R},\quad {\bm \nabla}q^*(\x)\cdot \n_0(\x) \rightarrow -\frac{\nu \owp \calB_a}{R_a},
\end{equation}
as $\x \rightarrow \x_a$
The solution is
\begin{equation}
\label{innerQ}
\fl Q_{a}(R)= 2  \owp \gamma_a^2\widehat{\lambda}_a  \calB_a +\calC_a \bigg [ \nu\ln( R /R_a)+\gamma_a \bigg ]-\nu^2 \owp \wlam_a\calB_a\bigg [ \ln R/R_a\bigg ] ^2
\end{equation}
for $R>R_a$ and the unknown coefficients $\calC_a$. The corresponding outer equation takes the form
\numparts
\begin{eqnarray}
\label{outerQ}
 {\fl \bm \nabla}^2 Q(\x)&=-2(2\pi  \nu  D_0)^2 \sum_{b,b'=1}^M\Gamma_{bb'} {\bm \nabla} G_{\Omega}(\x,\x_b)\cdot {\bm \nabla} G_{\Omega}(\x,\x_{b'})\nonumber\\
 \fl &=-(2\pi  \nu  D_0)^2 \sum_{b,b'=1}^M\Gamma_{bb'} {\bm \nabla}^2\bigg [ G_{\Omega}(\x,\x_b)G_{\Omega}(\x,\x_{b'})\bigg ]
\end{eqnarray}
for $\x\in \Omega\backslash \{\x_1,\ldots,\x_M\}$, with $ Q(\y)=0$ for all  $ \y \in \partial \Omega$ and the matching condition
\begin{eqnarray}
Q(\x)& \sim 2   \gamma_a^2\Gamma_{aa} + \calC_a\bigg [1+\gamma_a+ \nu  \ln (|\x-\x_a|/R_a) \bigg ]\nonumber \\
&\quad -  \Gamma_{aa}\bigg [1+ \nu  \ln (|\x-\x_a|/R_a) \bigg ]^2
\label{matchQ}
\end{eqnarray}
\endnumparts
as $\x\rightarrow \x_a$. We have also set 
\begin{equation}
\label{gamdef}
\Gamma_{ab}=\owp \calB_a\wlam_b.
\end{equation}

Equation (\ref{outerQ}) shows that the outer solution can be decomposed as
\begin{eqnarray}
Q(\x)=- (2\pi  \nu  D_0)^2 \sum_{b,b'=1}^M\Gamma_{bb'}G_{\Omega}(\x,\x_b)G_{\Omega}(\x,\x_{b'})+{\calR}(\x),
\end{eqnarray}
with
\begin{equation}
{\bm \nabla}^2 \calR(\x)=0,\ \x \notin \{\x_a,\, a=1,\ldots,M\}, \quad \calR(\y)=0,\ \y \in \partial \Omega.
\end{equation}
We now make the ansatz 
\begin{equation}
\label{outerR}
  \calR(\x)\sim -2\pi \nu  D_0 \sum_{b=1}^M  G_{\Omega}(\x,\x_b) \Delta_b
  \end{equation}
and match with the inner solution. The terms in $\ln^2|\x-\x_a|$ cancel. On the other hand, collecting terms in $\ln(|\x-\x_a|)$ and terms independent of $\ln(|\x-\x_a|)$, respectively, leads to the pair of equations
\begin{eqnarray}
& \calC_a-2\Gamma_{aa}\bigg [1- \nu \ln R_a \bigg ] 
  =\Delta_a+2\pi \nu  D_0\sum_{b=1}^M\bigg [ \Gamma_{ab} +\Gamma_{ba} \bigg ]\calG_{ab}.
\label{della}
\end{eqnarray}
and 
\begin{eqnarray}
&2\gamma_a^2\Gamma_{aa}+ \calC_a\bigg [1+\gamma_a - \nu \ln R_a \bigg ]-\Gamma_{aa}\bigg [1- \nu \ln R_a \bigg ]^2\nonumber \\
\fl &=-2\pi \nu D_0\sum_{b}\calG_{ab}\Delta_b- (2\pi \nu D_0)^2 \sum_{b,b'=1}^M \Gamma_{bb'}\calG_{ab}  \calG_{ab'}  .
\end{eqnarray} 
We have set $\calM_{ab}=2\pi D_0\calG_{ab}-\delta_{a,b}\ln R_a$.
Using equation (\ref{della}) to eliminate $\Delta_a$ and setting $\Gamma_{ab}=\owp \calB_a\wlam_b$ then gives
\begin{eqnarray}
 & \sum_{b=1}^M\bigg [{ \bm I}(\kappa_0)+\nu {\bm   \calM}\bigg ]_{ab} \calC_b=\owp \sum_{b=1}^M\Psi_{ab}(\kappa_0,\nu)\wlam_b,
 \label{Phiexp2}
 \end{eqnarray}
 where
  \begin{eqnarray}
\fl  & \Psi_{ab}(\kappa_0,\nu) 
  =(1-2\gamma^2_a(\kappa_0))\calB_a(\kappa_0)\delta_{a,b}+ 2\nu    \calM_{ab}\calB_b(\kappa_0)-\nu^2 K_{ab}(\kappa_0),
\label{Saab}
 \end{eqnarray}
 and
 \begin{eqnarray}
\fl  K_{ab}(\kappa_0)=\sum_{b,b'=1}^M  \bigg [\calM_{ab}  \calM_{ab'} -\calM_{ab}\calM_{bb'}-\calM_{ab'}\calM_{b'b}\bigg ]\calB_{b'}(\kappa_0) .
\end{eqnarray}

\subsection{Current fluctuations}

In summary, equation (\ref{Phiexp2}) relates the vector of effective Lagrange multipliers $\widehat{\bm \lambda}$ to the unknown vector of coefficients ${\bm \calC}$. We now substitute the inner solution $Q(\y)=Q_a(R_a)$ for all $\y \in \partial  \calU_a$ into equation (\ref{Fmulti}) with $d=2$:
 \begin{equation}
\label{Fmulti2}
Q_a(R_a)\equiv  2  \owp \gamma_a^2\widehat{\lambda}_a  \calB_a +\gamma_a\calC_a=\frac{ \delta J_a}{2\pi R_a  \kappa_0}.
\end{equation}
Combining with equation (\ref{Phiexp2}) generates a $\nu$-dependent linear relationship between the effective Lagrange multipliers and the surface fluctuations:
\begin{eqnarray}
 & \frac{1}{2\pi  \owp  D_0} \sum_{b=1}^M [{\bm I}(\kappa_0) +\nu {\bm \calM}]_{ab}\,  \delta \calJ_b\nonumber  \\
 &=2 \sum_{b=1}^M [{\bm I}(\kappa_0) +\nu {\bm \calM}]_{ab}\gamma_b(\kappa_0)\calB_b(\kappa_0)\wlam_b(\kappa_0)+ \sum_{b=1}^M\Psi_{ab}(\kappa_0,\nu)\wlam_b(\kappa_0)\nonumber \\
 &
\equiv  \sum_{b=1}^M\Phi_{ab}(\kappa_0,\nu)\wlam_b(\kappa_0) .\label{lamflu}
 \end{eqnarray}
Using a multi-target version of equation (\ref{S1target}) to evaluate the action along the optimal path, we have
\begin{eqnarray}
  S &\simeq \frac{1}{2} \sum_{a=1}^M \lambda_a \delta \calJ_a\nonumber \\
  &=\frac{1}{2} \bigg \{  {\bm \lambda}^{\top} \bigg  [{\bm I}(\kappa_0) +\nu {\bm \calM}\bigg ]^{-1}\bigg  [{\bm I}(\kappa_0) +\nu {\bm \calM} \bigg ] \delta  {\bm \calJ}  \bigg \} \nonumber \\
   &= \frac{1}{4\pi \owp  D_0}\delta\widehat {\bm  \calJ}^{\top}  {\bm \Phi}^{-1}(\kappa_0,\nu) \delta \widehat{\bm \calJ},
   \label{Snut}
\end{eqnarray}
where
\begin{equation}
 \delta \widehat{\bm\calJ}=\bigg  [{\bm I}(\kappa_0) +\nu {\bm \calM}  \bigg ]\delta {\bm  \calJ}.
\end{equation}
The corresponding multi-target distribution is
\begin{equation}
\calP[\delta{\bm \calJ}(T)=\delta {\bm \calJ},T]\asymp \exp\left (-\frac{T}{4\pi \owp  D_0}\delta\widehat {\bm  \calJ}^{\top} {\bm \Phi}^{-1}(\kappa_0,\nu) \delta \widehat{\bm \calJ}\right ).
\end{equation}
It immediately follows that $\langle\delta \widehat{\calJ}_a(T)\rangle =0$ and
\begin{equation}
\bigg \langle \delta \widehat{\calJ}_a(T)  \delta \widehat{\calJ}_b(T)\bigg \rangle =\frac{2\pi \owp  D_0}{T} {\Phi}_{ab}(\kappa_0,\nu).
\label{cov}
\end{equation}
In summary, we have derived equation (\ref{mLDP}) with 
\begin{equation}
\fl {\bm \Sigma}(\kappa_0,\nu) =2\pi \owp  D_0\bigg  [{\bm I}(\kappa_0) +\nu {\bm \calM}  \bigg ]^{-1}{\bm \Phi}(\kappa_0,\nu)  \bigg  [{\bm I}(\kappa_0) +\nu {\bm \calM}  \bigg ]^{-1}.
\end{equation}
Moreover, the covariance matrix is a non-perturbative function of $\nu$.

\subsubsection*{Single target.} In the case of a single target, the matrix equations (\ref{AP}) and (\ref{Phiexp2}) reduce to scalar equations. In particular, equations (\ref{Phiexp2}) and (\ref{Fmulti2}) then become, respectively,
\begin{eqnarray}
\calC_1= \owp \lambda_1\Psi_1(\nu),\quad \Psi_1(\nu)
\equiv   \frac{1-2\gamma_1^2+2\nu \calM_{11} +\nu^2 \calM_{11}^2}{(1+\gamma_1+\nu \calM_{11})^3} ,
\end{eqnarray}
and
 \begin{equation}
\frac{ \delta J_1}{2\pi \owp D}
=   \lambda_1\bigg (\frac{2\gamma_1}{(1+\gamma_1+\nu \calM_{11})^2}  +\Psi_1(\nu)\bigg ).
\end{equation}
The corresponding action is
\begin{equation}
S=\frac{\lambda_1 \delta \calJ_1}{2} \sim \bigg (\frac{2\gamma_1}{(1+\gamma_1+\nu \calM_{11})^2}  +\Psi_1(\nu)\bigg )^{-1}\frac{ \delta J^2_1}{4\pi \owp D},
\end{equation}
and the variance is
\begin{equation}
 \langle\delta J_1^2\rangle \sim \frac{2\pi \owp D}{T} \bigg (\frac{2\gamma_1}{(1+\gamma_1+\nu \calM_{11})^2}  +\Psi_1(\nu)\bigg ).
\end{equation}

\subsubsection*{Small-$\nu$ limit for multiple targets.} In the case of multiple targets, one can avoid having to perform any matrix inversions by carrying out a regular perturbation expansion in powers of $\nu$. First, expanding equation (\ref{Fmatrix}) to $O(\nu)$ gives
\begin{equation}
 \calB_a\sim  \frac{1}{1+\gamma_a}\bigg (1+\nu \sum_b\calM_{ab}(1+\gamma_b)^{-1}+O(\nu^2)\bigg ).
 \end{equation}
 Hence, expanding equation (\ref{Saab}) and substituting into equation (\ref{lamflu}), we have
\begin{eqnarray}
\fl \Phi_{ab}(\nu) 
&  \sim  \frac{2\gamma_a+1}{1+\gamma_a}\bigg (1+\nu \sum_c\calM_{ac}(1+\gamma_c)^{-1}\bigg ) \delta_{a,b}+ 2\nu  \calM_{ab}  +O(\nu^2).
\label{Saab3}
 \end{eqnarray}
Moreover
\begin{eqnarray}
\label{haa}
\fl & \bigg \langle \delta \widehat{\calJ}_a(T)  \delta \widehat{\calJ}_b(T)\bigg \rangle \\
\fl &=\bigg \langle \bigg [(1+\gamma_a)\delta {\calJ}_a(T) +\nu \sum_{a'=1}^M \calM_{aa'}\delta \widehat{\calJ}_{a'}(T)\bigg ]\bigg [(1+\gamma_b)\delta {\calJ}_b(T) +\nu \sum_{b'=1}^M \calM_{ba'}\delta \widehat{\calJ}_{a'}(T)\bigg ]\bigg \rangle \nonumber \\
\fl &=(1+\gamma_a)(1+\gamma_b)\frac{\Sigma_{ab} }{T}+\nu \sum_{a'}\bigg [ (1+\gamma_b)\calM_{aa'} \frac{\Sigma_{a'b}}{T} +(1+\gamma_a)\calM_{ba'}\frac{\Sigma_{aa'}}{T}\bigg ],\nonumber \end{eqnarray}
where
\begin{equation}
\frac{\Sigma_{ab}}{T}=\bigg \langle \delta {\calJ}_a(T)  \delta {\calJ}_b(T)\bigg \rangle .
\end{equation}
Substituting the series expansion
$\Sigma_{ab}= \Sigma_{ab}^{(0)}+\nu\Sigma_{ab}^{(1)}+\ldots $
into equation (\ref{haa}) and combining with equations (\ref{cov}) and (\ref{Saab3}), we find that
\begin{equation}
\label{var0}
\Sigma_{ab}^{(0)} = 2\pi \owp  D_0 \frac{1+2\gamma_a}{(1+\gamma_a)^3} \delta_{a,b},
\end{equation}
and
\begin{eqnarray}
\fl   2\pi \owp  D_0  (1+\gamma_a )(1+\gamma_b) \Sigma_{ab}^{(1)}&=\calM_{ab} \bigg [2- \frac{1+2\gamma_b}{(1+\gamma_b)^2} - \frac{1+2\gamma_a}{(1+\gamma_a)^2}\bigg ]\nonumber \\
\fl  &\quad + \frac{2\gamma_a+1}{1+\gamma_a} \bigg ( \sum_c\calM_{ac}(1+\gamma_c)^{-1} \bigg ) \delta_{a,b}.
\label{var1}
\end{eqnarray}
First note that the expression for $\Sigma_{aa}^{(0)}$ recovers the variance (\ref{varran}) for a single target in an annular domain with inner radius $R=\epsilon L$ such that $\nu=1/\ln (L/R)$. This follows from rewriting $\gamma_a$ in unscaled parameters,
\begin{equation*}
\gamma_a=\frac{\nu D}{\kappa R} =\frac{ D}{\kappa R}\frac{1}{\ln(L/R)},
\end{equation*}
so that 
\begin{eqnarray*}
\fl \frac{1+2\gamma_a}{(1+\gamma_a)^3} \nu D =\frac{1+\frac{\displaystyle 2D}{\displaystyle \kappa R\ln (L/R)}}{\displaystyle \bigg (1+\frac{2D}{\kappa R\ln (L/R)}\bigg )^3} \frac{D}{\ln (L/R)}=\frac{\ln (L/R)+\frac{\displaystyle 2D}{\displaystyle \kappa R}}{\displaystyle \bigg (\ln (L/R)+\frac{2D}{\kappa R }\bigg )^3}D \ln(L/R).\nonumber \\
\end{eqnarray*}
Second, off-diagonal terms in the covariance matrix occur at $O(\nu)$, which correspond to cross-correlations in the current fluctuations. However, taking the totally absorbing limit $\kappa_0\rightarrow \infty$ of equation (\ref{var1}) we see that the foff-diagonal term on the right-hand side vanishes since $\gamma_a\rightarrow 0$ for all $a=1,\ldots,M$. We then recover the  result obtained in Ref. \cite{Agranov17}.

\begin{figure}[t!]
\centering
\includegraphics[width=5.5cm]{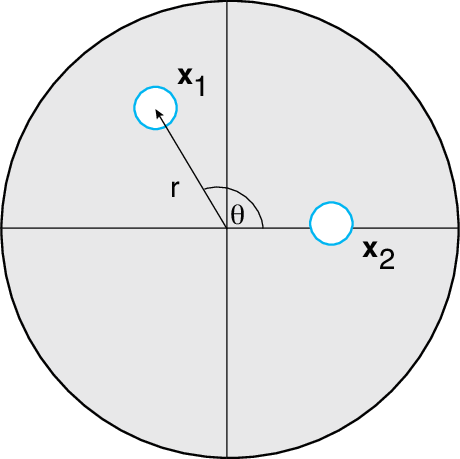} 
\caption{A pair of targets in the unit disc with centres $\x_1=r(\cos \theta,\sin \theta)$ and $\x_2=(x_2,0)$. Each has a radius $\epsilon$, $0<\epsilon \ll 1$.}
\label{fig5}
\end{figure}

\begin{figure}[t!]
\centering
\includegraphics[width=10cm]{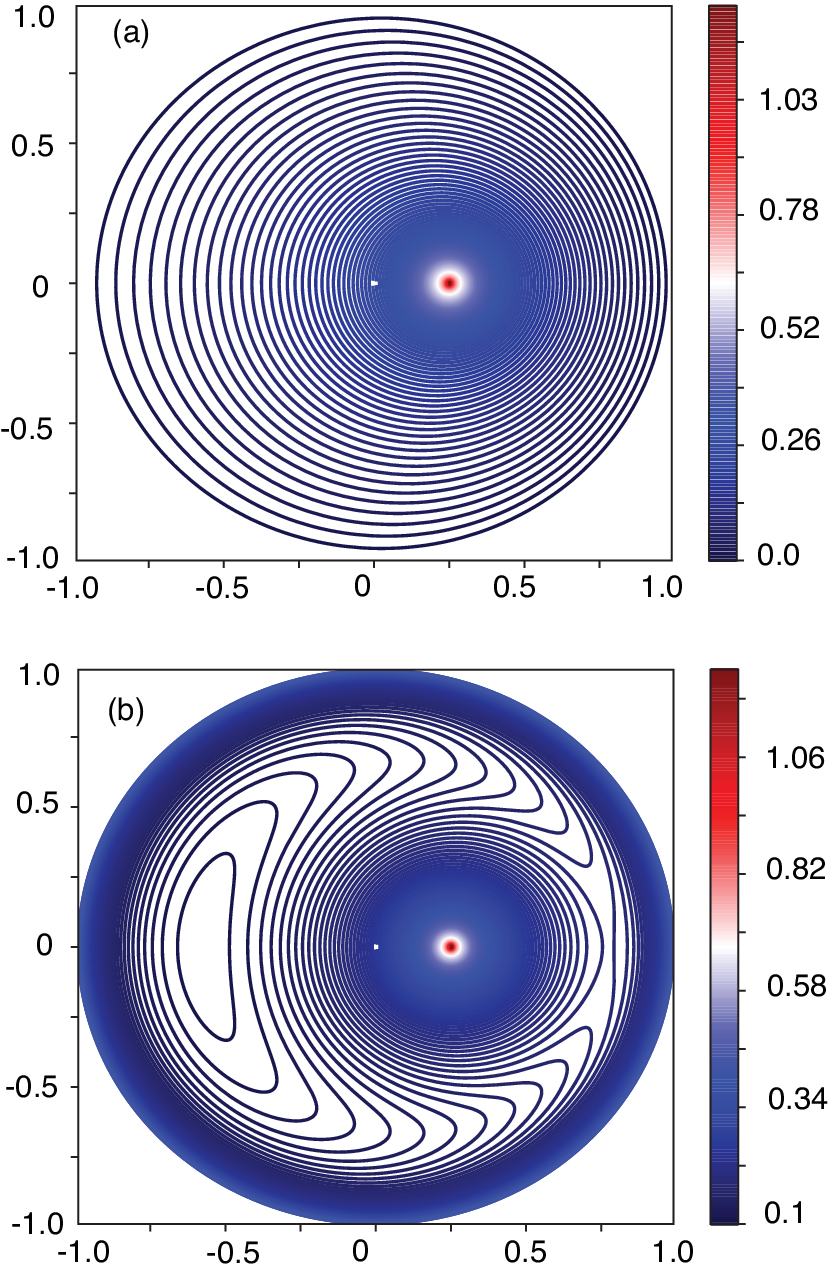} 
\caption{Contour plots of the $O(\nu)$ variance $ \Sigma_{aa}^{(1)}(\nu)$ for the pair of targets shown in Fig. \ref{fig5} as a function of the position $\x_2$ of the second target expressed in polar coordinates. (a) First target ($ \Sigma_{11}^{(1)}(\nu)$). (b) Second target ($ \Sigma_{22}^{(1)}(\nu)$). We take $|\x_2|=0,25$, $2\pi \owp D_0=1$ and $\gamma_0=1$.}
\label{fig6}
\end{figure}

\subsubsection*{Pair of targets in the the unit disc.}
\begin{figure}[t!]
\centering
\includegraphics[width=13cm]{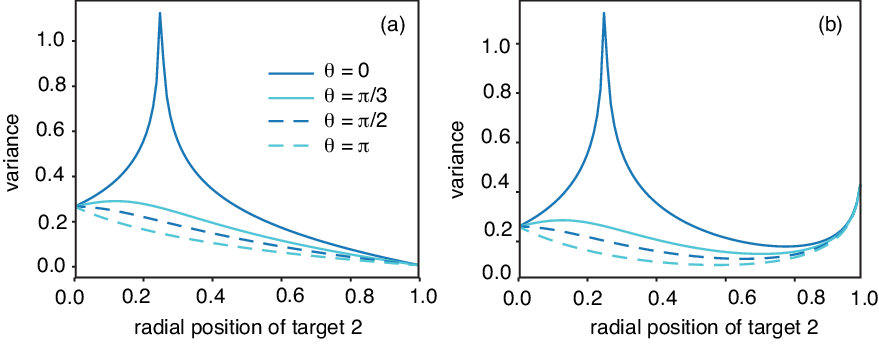} 
\caption{Plots of the leading-order variances (a) $\Sigma^{(1)}_{11}$ and (b) $\Sigma^{(1)}_{22}$ as a a function of the radial position of target 2 for different angular positions $\theta$. Same setup as Fig. \ref{fig6}.}
\label{fig7}
\end{figure}

In order to illustrate the above perturbation calculations, consider the 2D configuration shown in Fig. \ref{fig5}. The domain $\Omega$ is taken to be the unit disc, which contains a target of radius $\epsilon$ placed at $\x_{1}=r(\cos \theta,\sin\theta)$ and a second target at $\x_2=(x_2,0)$. The full Dirichlet Green's function is
\begin{eqnarray}
\fl G_{\Omega}(\x,\hxi)&=\frac{1}{2\pi}\bigg[-\ln|\x-\hxi|+\ln   |\x- \hxi^*|+\ln|\hxi|   \bigg ],\quad \hxi^*=\frac{\hxi}{|\hxi |^2},
\label{Gdisc}
\end{eqnarray}
with the regular part obtained by dropping the first logarithmic term.
Applying equations (\ref{var0}) and (\ref{var1}) to the two-target case with $\gamma_a=\gamma_0\equiv D_0/\kappa_0$ for $a=1,2$, we have 
\begin{equation}
\fl \Sigma_{aa}^{(1)}=  2\pi \owp  D_0  \bigg (\frac{2}{(1+\gamma_0)^2}M_{aa}+\frac{1+2\gamma_0}{(1+\gamma_0)^4}\sum_{b=1,2} \bigg [ \calM_{ab} -\calM_{aa} \bigg ]\bigg ),
\end{equation}
and
\begin{eqnarray}
 \Sigma_{12}^{(1)}=  4\pi \owp  D_0\bigg [1- \frac{1+2\gamma_0}{(1+\gamma_0)^2}\bigg ] \frac{\calM_{12} }{(1+\gamma_0)^2}.
\end{eqnarray}
In Fig. \ref{fig6} we show contour plots of the $O(\nu)$ variance contributions $\Sigma_{11}^{(1)}$ and $\Sigma_{22}^{(1)}$ as functions of the second target's position in polar coordinates. (The corresponding plot of the covariance $\Sigma_{12}^{(1)}$  is identical to $\Sigma_{11}^{(1)}$ (up to a scale factor.) The hotspot around the position of the first target reflects the singular nature of the Green's function and the breakdown of the asymptotic analysis if the distance between the two targets is $O(\epsilon)$. The analysis also breaks down if the first target is within an $O(\epsilon)$ distance of the unit circle $\partial \Omega$. This is further illustrated in Fig. \ref{fig7}. 

\section{Discussion}

In this paper we analysed current fluctuations in a non-interacting Brownian gas with one or more partially absorbing targets within a bounded domain $\Omega \subset \R^d$. Starting from the microscopic theory of Brownian particles interacting with a partially reactive surface $\partial \calU$, we derived a DK equation with a Robin boundary condition on $\partial \calU$, see equations (\ref{RBDK1}) and (\ref{RBDK2}),  under the mean field ansatz (\ref{ans}). Coarse-graining the DK equation, we derived MFT equations for the optimal noise-induced path for a single partially reactive target under a saddle-point approximation of the associated path integral action. We then obtained a Gaussian probability distribution for typical current fluctuations by linearising the MFT equations about the corresponding deterministic or noise-averaged system and solving the resulting stationary equations. We illustrated the theory by considering two simple geometric configurations, namely,  the finite interval and a circular annulus. Finally, we extended our analysis to multiple partially absorbing targets. First, we derived a multi--target  Gaussian probability distribution for the currents in the case of an arbitrary geometric configuration, and used this to show that there are cross-correlations in the statistics of current fluctuations for partially absorbing targets, even though the  Brownian gas is non-interacting. We then explicitly calculated the covariance matrix for circular targets in a 2D bounded domain $\Omega$ of characteristic size $L$. The targets were assumed to have radii of $O(\epsilon L)$, $0<\epsilon \ll 1$, and to be well-separated. We used singular perturbation theory to derive a non-perturbative expression for the covariance matrix with respect to the small parameter $\nu=-\ln \epsilon$. We found that the leading order contribution is a diagonal matrix whose entries are consistent with the result for a circular annulus, whereas higher-order contributions have off-diagonal terms and are non-local functions of the target positions and radii. There are a number of issues worth exploring in future work.

\noindent {\em (i) Fully stochastic DK equations and the mean field ansatz.} Our derivation of the generalised DK equation (\ref{RBDK1}) with the Robin boundary condition (\ref{RBDK2}) was based on the assumption that at the microscopic level, an individual particle is absorbed when its local time crosses an exponentially distributed threshold $h_j$, see section 2. In particular, equations (\ref{RBDK1}) and (\ref{RBDK2}) followed from
averaging  the empirical density $\wp(\x,\h,t)$ and the flux $\nu(\x,\h,t)$ with respect to $\h$ in equations (\ref{toto1}) and (\ref{toto2}), and imposing the mean field ansatz (\ref{ans}). Two open problems are (i) determining the validity of the mean field ansatz and (ii) developing MFT for the fully stochastic DK equations. In the latter case, there are then two major challenges that need to be confronted. First, equations (\ref{toto1}) and (\ref{toto2}) do not define a closed BVP for the stochastic density $\wp(\x,\h,t)$. (A similar comment holds for the analog of the partially averaged system given by equations (\ref{RBDK1}) and (\ref{RBDK2}) in the case of a non-exponential density $\psi(h)$.) Second, if an appropriate path integral representation of sample paths could be constructed, then how would this be modified by averaging with respect to the local time thresholds? Following Ref. \cite{Derrida09}, this would be an example of {\em annealed averaging}. Moreover, in the case of exponentially distributed thresholds, how would
annealed averaging differ from averaging the empirical density prior to constructing MFT, which was the approach taken in the current paper?
\medskip

\noindent {\em  (ii) Interacting particle systems.} In this paper, we restricted our analysis to non-interacting Brownian particles. The mapping from the microscopic to the macroscopic theory was achieved by coarse-graining the generalised DK equations (\ref{RBDK1}) and (\ref{RBDK2}), which were obtained by modelling absorption in terms of quenched local time thresholds with exponential distributions. One natural way to include interactions would be to consider a pairwise potential $V(\X_j-\X_k)$, where $\X_j$ and $\X_k$ are the positions of the interacting pair. The SDE (\ref{Sko}) becomes
\begin{eqnarray}
\fl  d\X_j(t)&=-\frac{1}{\gamma }\sum_{k=1}^N  V'(\X_j(t)-\X_k(t)) dt +\sqrt{2D}d{{\bf W}}_j(t)-{\bm n}_0(\X_j(t))dL_j(t),
 \label{abs0}
 \end{eqnarray}
 where $\gamma$ is a friction coefficient. It is relatively straightforward to extend the derivation of the generalised DK equation in section 2 to include pairwise interactions. The result for a quenched local time threshold is, see equations (\ref{toto1}) and (\ref{toto2}),
  \numparts
\begin{eqnarray} 
\label{intRB1}
 &\frac{\partial \wp(\x,\h,t)}{\partial t} =-{\bm \nabla} {\bm \frJ}(\x,\h,t) ,\quad \x\in \R^d\backslash \calU,\\
& {\bm \frJ}(\y,\h,t)\cdot \n_0(\y)=D\nu(\y,\h,t),\quad \y\in \partial \calU,
\label{intRB2}
\end{eqnarray}
\endnumparts
with
\begin{eqnarray}
\label{intRB3}
 \fl {\bm \frJ}(\x,\h,t) &=   \sqrt{2D}\sum_{j=1}^{N_0}  \wp_{j}(\x,\h,t) {\bm \xi}_j(t) -D{\bm \nabla}\wp(\x,\h,t)-{\bm \frK}(\x,\h,t),
\end{eqnarray}
and
\begin{eqnarray}
{\bm \frK}(\x,\h,t)&= \gamma^{-1} \wp(\x,\h,t)  \int_{\R^d\backslash \calU} d\z\,{\wp}(\z,\h,t){\bm \nabla}V(\x-\z).
\end{eqnarray}
Averaging with respect to the Gaussian white noise processes in the presence of pairwise particle interactions results in a well-known moment closure problem \cite{Dean96,Bressloff24}. That is, the average field $\rho(\x,\h,t)=\langle \wp(\x,\h,t)\rangle$ couples to the two point-correlation field $\langle \wp(\x,\h,t)\wp(\y,\h,t)\rangle$ etc. An alternative to constructing a DK equation for interacting Brownian particles would be to consider 
 the macroscopic limit of a lattice gas model. Indeed MFT was originally developed within the context of lattice gas models \cite{Bertini15}, whose density equations have the general form given by equations (\ref{lattice1}) and (\ref{lattice2}). First, however, it would be necessary to incorporate a microscopic theory of partially reactive targets analogous to the derivation of the generalised DK equation.  Another challenge is that even for totally absorbing targets, it is more difficult to obtain exact solutions of the linearised MFT equations \cite{Agranov17}.
\medskip

\noindent {\em (iii) Collective local times.}  Another possible scenario that lends itself to MFT is to assume that each contact between a particle and the boundary modifies an internal state of the boundary such that the latter switches from a totally reflecting to a totally absorbing surface, say, when the collective local time $L(t)=\sum_{j=1}^NL_j(t)$ exceeds a threshold $h$. One potential quantity of interest is the stopping time for such a surface transition,
\begin{equation}
\calT=\inf\{t>0 , L(t)>h\}.
\end{equation}
The statistics of $\calT$ will then depend on the statistics of the collective local time $L(t)$. Note that MFT has recently been used to analyse the latter in the case of a non-interacting Brownian gas in $\R$, but without any connection to absorption processes \cite{Smith24}. It would be interesting to adapt this analysis to analyse the statistics of $\calT$ on the half-line (or a higher-dimensional domain).

\setcounter{equation}{0}
\renewcommand{\theequation}{A.\arabic{equation}}
\section*{Appendix A: Derivation of Euler-Lagrange equations}

Consider the probability that $q(\x,T)=q_1(\x)$ given that $q(\x,0)=q_0(x)$. Formally speaking, after discretizing time we have
\begin{eqnarray*}
\fl &\P[\{q_1(\x)\}|\{q_0(\x), {\bf W}(\x)\}]\nonumber \\
\fl &=\int D[q,j]\prod_{\x} \prod_n \delta(q_{n+1}(\x)-q_n(\x)-D { \bm \nabla}\cdot {\bm j}_n(\x)\Delta t)\\
\fl &\qquad \times \delta([{\bf j}_{n}(\x)+D{\bm \nabla} q_n(\x)]\Delta t -\sqrt{2\epsilon D q_n(\x))}\Delta {\bf W}_n(\x)),
\end{eqnarray*}
where $\epsilon =T^{-d/2}$ and $\Delta {\bm W}_n(\x)={\bm W}_{n+1}(\x)-{\bm W}_n(\x)={\bm \eta}_n(\x)\Delta t$.
Using the Fourier representation of the second Dirac delta function yields
\begin{equation}
 \int_{\R^d} \e^{\displaystyle i{\bm \uhat}_{m}\cdot \left ([{\bm j}_{m}+D { \bm \nabla}q_m]\Delta t-{\sqrt{2\epsilon D q_m}}\Delta \W_{m}\right ) }\frac{\displaystyle d{\bm \uhat}_{m}}{\displaystyle  (2\pi)^d}.
\end{equation}
Integrating with respect to the Gaussian noise gives
\begin{equation}
 \int_{\R^d} \e^{i{\bm \uhat}_{m}\cdot \left ({\bm j}_{m}+D {\partial_x}{\bm \nabla} q_m\right )\Delta t}\e^{-2\epsilon Dq_m{\bm \uhat}_m^2 \Delta t/2}\frac{d{\bm \uhat}_{m}}{(2\pi)^d},
\end{equation}
and then performing the Gaussian integral with respect to ${\bm \uhat}_{m}$ yields
\begin{eqnarray}
\frac{1}{\sqrt{4\pi \epsilon Dq_m \Delta t}}\e^{-\Delta t\left   ({\bm j}_{m}+D{ \bm \nabla}q_m\right )^2/(4\epsilon D q_m}.
\end{eqnarray}
Retaking the continuum time limit and absorbing various factors into the measure yields the path integral representation (\ref{pathint}).

We calculate the first variation of the corresponding action appearing in equation (\ref{Seff}) as follows
\begin{eqnarray}
\fl\delta \calS&=\  \int_0^1 dt\, \int_{\Omega\backslash \calU}d\x\, \bigg \{ \frac{[{\bm j}+D{\bm \nabla}q ]\cdot [\delta {\bm j} +D{\bm \nabla}\delta q ]}{2Dq}\nonumber \\
\fl &\hspace{3cm} -\frac{[{\bm j}+D{\bm \nabla}q]^2}{4Dq ^2}\delta q+\delta p[\partial_tq+{\bm \nabla}\cdot {\bm j}]+p[\partial_t\delta q+{\bm \nabla}\cdot \delta {\bm j}] \bigg \}\nonumber \\
\fl &\hspace{2cm} -\lambda  \int_0^1 dt\,  \int_{\partial \calU}d\y\, \delta {\bm j}\cdot \n_0     .
\end{eqnarray}
Integrating by parts all terms involving derivatives of $\delta q,\delta {\bm j}$ and rearranging gives
\begin{eqnarray}
\fl\delta \calS&= \int_0^1 dt\, \int_{\Omega\backslash \calU}d\x\,\left \{ \delta {\bm j}\cdot  \left (\frac{{\bm j}+D{\bm \nabla}q }{2Dq}-{\bm \nabla p}\right ) +\delta p\bigg (\partial_tq+{\bm \nabla}\cdot {\bm j}\bigg )\right \}\nonumber \\
\fl & \quad -  \int_0^1 dt\, \int_{\Omega\backslash \calU}d\x\, \delta q\left \{  \left ( \frac{[{\bm j}+D{\bm \nabla}q]^2}{4Dq ^2}\right )+{\bm \nabla}\cdot \frac{{\bm j}+D{\bm \nabla}q}{2Dq}+\partial_t p\right \}\nonumber \\
\fl & \quad +\int_0^1 dt\,  \int_{\partial \calU}d\y\,  \frac{[{\bm j}+D{\bm \nabla}q ]\cdot \n_o}{2q} \delta q + \int_0^1 dt\,  \int_{\partial \calU}d\y\, \bigg (p-\lambda\bigg ) \delta {\bm j} \cdot \n_0
\label{varvar}  \\
\fl & \quad +\int_0^1 dt\,  \int_{\partial \Omega}d\y\,  \frac{[{\bm j}+D{\bm \nabla}q ]\cdot \n_o}{2q} \delta q + \int_0^1 dt\,  \int_{\partial \Omega}d\y\,  p \,\delta {\bm j} \cdot \n_0   +  \int_{\Omega\backslash \calU}d\x \, p  \delta q |_{t=1}  .\nonumber
\end{eqnarray}
The free variation of $\delta {\bm j}$ in $\Omega\backslash \calU$ requires ${\bm j}+D{\bm \nabla}q=2Dq{\bm \nabla p}$, whereas the corresponding free variations of $\delta p$ and $\delta q$ yield equations (\ref{HD1}) and (\ref{HD2}), respectively. The Robin boundary condition on $\partial \calU$ implies that $\delta {\bm j} \cdot \n_0 =-\kappa_0 \delta q$. Applying this constraint to the third line of equation (\ref{varvar}) results in the boundary condition (\ref{HD3}). Similarly, the boundary condition on $\partial \Omega$ means that $\delta q=0$ on $\partial \Omega$. and, hence, the fourth line imposes the boundary condition $p(\y,t)=0$ on $\partial \Omega$. Finally, the last term on the right-hand side of equation (\ref{varvar}) yields the condition $p(\x,t=1)=0$.

\setcounter{equation}{0}
\renewcommand{\theequation}{B.\arabic{equation}}
\section*{Appendix B: Potential theory for multiple targets: derivation of equation (\ref{nabka2})}

Following along similar lines to the spectral analysis in section 3.2, the solution of equations (\ref{pot1})-(\ref{pot3}) for the potential $\phi_{\ell}$ is
\begin{eqnarray}
 \label{mnab}
\phi_{\ell}(\x)=-  D\sum_{a=1}^M\int_{\partial \calU_a}d\z\, \Phi_{\ell}(\z){\bm \nabla}_{\z}G(\z,\x)\cdot \n_{a}(\z) .
\end{eqnarray}
where $G$ is the Dichlet Green's function on $ \Omega \backslash \calU$ with $\calU=\cup_{a=1}^M  \calU_a$, and $\phi_{\ell}(\y)=\Phi_{\ell}(\y)$ for all $\y \in \partial \calU$. Imposing the boundary condition (\ref{pot2}) then yields the integro-differential equation
 \begin{equation}
\label{Phi}
  {\mathbb L }[\Phi_{\ell}](\y)+\frac{\kappa}{D}[\Phi_{\ell}(\y)-\delta_{a,\ell}]=0,\quad \y \in \partial \calU_{a},
\end{equation}
where ${\mathbb L} $ is the D-to-N operator (\ref{DtoN2}) acting on the space $L^2(\partial \calU)$.
Denoting the eigenvalues and eigenfunctions of ${\mathbb L}$ by $\mu_n$ and $v_n(\y)$, $\y \in \partial \calU$, and introducing the eigenfunction expansion
\begin{equation}
\label{mPhi2}
\Phi_{\ell}(\y)=\sum_{m=0}^{\infty}\Phi_{\ell,m } v_m(\y).
\end{equation}
we find that
 \begin{equation}
\label{mPhm}
\bigg (\mu_n+\frac{\kappa}{D}\bigg )\Phi_{\ell,n}=\frac{\kappa }{D} \overline{v}_{\ell,n},\quad \overline{v}_{\ell,n}=\int_{\partial \calU_{\ell}}v_n(\y)d\y.
\end{equation}
Combining with equation (\ref{mnab}) shows that
\begin{equation}
\phi_{\ell}(\x)=\sum_{n}\frac{\kappa/D}{\mu_{n}+\kappa/D}\overline{v}_{\ell,n}\phi_n(\x),
\label{mpixie1}
\end{equation}
and
\begin{equation}
 \phi_n(\x)=-D\sum_{a=1}^M \int_{\partial \calU_a} d\y\, v_n(\y) {\bm \nabla} _{\y}G(\y,\x)\cdot \n_a(\y).
 \label{boil}
\end{equation}

Next, replacing the boundary condition (\ref{HDlin2}) by the Dirichlet boundary condition
$Q(\y)=\Psi(\y)$ for $ \y \in \partial \calU$
we otain the formal solution of equation (\ref{HDlin3}):
\begin{eqnarray}
  Q(\x)&= -2D  \int_{\Omega\backslash \calU}d\z\, G(\x,\z){\bm \nabla}\cdot \bigg [q^*(\z) {\bm \nabla}P(\z)\bigg ] \nonumber \\
  &\quad -D\sum_{a=1}^M \int_{\partial \calU_a}d\z\,  \Psi(\z){\bm \nabla}_{\z} G(\z,\x)\cdot \n_a(\z) .
\end{eqnarray}
Imposing the boundary condition (\ref{HDlin4}) then yields the integro-differential equation
\begin{eqnarray}
\label{Lin00}
\fl   {\mathbb L}[\Psi](\y)+\frac{\kappa}{D} \Psi(\y) &=-2D   \int_{\Omega\backslash \calU}d\z\, \bigg ( {\bm \nabla}_{\y}G(\y,\z)\cdot \n_a(\y)\bigg ){\bm \nabla}\cdot \bigg [q^*(\z) {\bm \nabla}P(\z)\bigg ]\nonumber \\
\fl &\quad +2q^*(\y){\bm \nabla }P(\y)  \cdot \n_a(\y) ,\,\  y \in \partial \calU_a.
\end{eqnarray}
Introducing the eigenvalue expansion
\begin{equation}
\Psi(\y)=\sum_{n=0}^{\infty}\Psi_nv_n(\y),
\end{equation}
and using equation (\ref{boil}) we have 
\begin{eqnarray}
\bigg (\mu_n+\frac{\kappa}{D}\bigg )\Psi_{n}&=2 \int_{\Omega\backslash \calU}d\z\, \phi_n(\z){\bm \nabla}\cdot \bigg [q^*(\z) {\bm \nabla}P(\z)\bigg ]\nonumber \\
  &\quad  +2\sum_{a=1}^M\int_{\partial \calU_a}d\y\, v_n(\y)q^*(\y){\bm \nabla }P(\y)  \cdot \n_a(\y).
\end{eqnarray}
It follows from equation (\ref{Fmulti}) that
\begin{eqnarray}
 \fl  & T^{1-d/2}\delta \calJ_b \\
 \fl &\equiv  \kappa \int_{\partial \calU_b}\Psi(\y)d\y=\kappa\sum_{n}\Psi_n \overline{v}_{b,n} \nonumber \\
 \fl  &=-2D \int_{\Omega\backslash \calU}d\z\, \phi_b(\z){\bm \nabla}\cdot \bigg [q^*(\z) {\bm \nabla}P(\z)\bigg ] +2D\sum_{a=1}^M\int_{\partial \calU_a}d\y\, \Phi_b(\y)q^*(\y){\bm \nabla }P(\y)  \cdot \n_a(\y)\nonumber \\
 \fl  & =-2D \int_{\Omega\backslash \calU}d\z\, \phi_b(\z){\bm \nabla}\cdot \bigg [q^*(\z) {\bm \nabla}P(\z)\bigg ] +2D\int_{\Omega \backslash \calU}d\y\,{\bm \nabla}\cdot  \bigg [\phi_b(\z)q^*(\y){\bm \nabla }P(\y) \bigg ]\nonumber \\
 \fl&=2D\int_{\Omega \backslash \calU}d\x\, q^*(\x){\bm \nabla}\phi_b(\x) \cdot {\bm \nabla }P(\x)  
 =2D\sum_{a=1}^M\lambda_a \int_{\Omega \backslash \calU}d\x\, q^*(\x){\bm \nabla}\phi_b(\x) \cdot {\bm \nabla }\phi_a(\x) .\nonumber 
\end{eqnarray}
We thus obtain equation (\ref{nabka2}).

\end{document}